\documentclass[aps,reprint,showpacs,prb,superscriptaddress]{revtex4-1}

\usepackage{amsmath,amssymb,graphicx,units}
\usepackage[usenames,dvipsnames]{color}

\DeclareMathOperator{\Tr}{Tr}

\begin{document}

\title{Thermodynamics of two-dimensional spin models with bimodal
  random-bond disorder} \author{Baoming Tang} \affiliation{Department
  of Physics, The Pennsylvania State University, University Park, PA
  16802, USA} \affiliation{Department of Physics, Georgetown
  University, Washington, DC 20057, USA} \author{Deepak Iyer}
\author{Marcos Rigol} \affiliation{Department of Physics, The
  Pennsylvania State University, University Park, PA 16802, USA}

\begin{abstract}
  We use numerical linked cluster expansions to study thermodynamic
  properties of the two-dimensional classical Ising, quantum XY, and quantum Heisenberg
  models with bimodal random-bond disorder on the square and honeycomb
  lattices. In all cases, the nearest-neighbor coupling between the
  spins takes values $\pm J$ with equal probability.  We obtain the
  disorder averaged (over all disorder configurations) energy,
  entropy, specific heat, and uniform magnetic susceptibility in each
  case.  These results are compared with the corresponding ones in the
  clean models.  Analytic expressions are obtained for low orders in
  the expansion of these thermodynamic quantities in inverse
  temperature.
\end{abstract}


\maketitle

\section{Introduction}
\label{sec:intro}

Solid-state materials deviate in various ways from the periodic
idealizations sometimes used to describe them theoretically. In
crystals, for example, such deviations can occur due to the presence
of lattice distortions, impurity atoms (that may or may not be
magnetic), and vacancies in the lattice, collectively termed as
disorder.  Disorder can significantly influence material properties. A
dramatic example for noninteracting electrons is Anderson
localization.\cite{anderson58} In spin systems, the focus of this
work, quenched disorder in the spin interactions leads to frustration
and can generate spin glasses.\cite{binder86} A spin glass is a
remarkable state of matter where, loosely speaking, spins are
``frozen'' in an irregular pattern, i.e., they display a very slow
dynamics under external driving. Although this phase does not exhibit
spin order in a traditional sense (e.g., as a ferromagnet), it is
still distinctly different from a paramagnet. Experimentally, spin
glasses are generally associated with a cusp in the ac susceptibility
at a certain temperature above which the system behaves like a
paramagnet.\cite{weissman93} Whereas no standard order parameter
captures a glassy transition, glass order parameters exist that do so
(see, e.g., Refs.~\onlinecite{binder86,Katzgraber07}).

In two-dimensional (2D) lattices (the focus of this work), the effects
of quenched disorder on classical spins have been extensively studied
over the years.\cite{binder86} However, the calculation of
thermodynamic properties and correlation functions continues to be a
computational
challenge.\cite{parisen_francesco_11,thomas_middleton_13} While it is
generally believed that no spin-glass phase exists for nonzero
temperatures,\cite{binder86} zero-temperature phases continue to be
debated for various models of interest (see, e.g.,
Refs.~\onlinecite{thomas_huse_11,parisen_francesco_11}).  For {\it
  quantum} spin models, the sign problem\cite{henelius00,troyer05} in
quantum Monte Carlo simulations and the exponential growth of the
Hilbert space, relevant to full exact diagonalization calculations,
represent an even greater challenge. Because of this, the properties
of disordered quantum spin systems, and of quantum spin glasses in
particular, have remained essentially unexplored. The existing
literature in the subject has almost exclusively dealt with classical
models.

Our goal in this work is to use a recently introduced numerical linked
cluster expansion (NLCE) for disordered systems\cite{tang2014}
to study the thermodynamic properties of the classical Ising (with
$S=1/2$), and quantum (spin-1/2) XY and Heisenberg models with bimodal
random-bond disorder on the square and honeycomb lattices. NLCEs allow
us to obtain finite temperature properties of those models in the
thermodynamic limit through the exact diagonalization of finite-size
clusters.  We specifically study the energy, entropy, specific heat,
and uniform magnetic susceptibility (for the magnetization in the
$z$-direction) as a function of temperature. Since any glassy phase is
only expected to emerge at zero temperature in these models, if at
all, we do not study the spin-glass order parameter. In future work,
we will study quantum
quenches\cite{tang2014,rigol_14a,wouters_denardis_14,rigol_14b} to
examine the possibility of disorder driven localization in 2D.

Our results are briefly as follows: For clean systems, they match well
with known results for the square lattice, while we report additional results
for the honeycomb lattice.  For the disordered systems, we unveil some
interesting features. In the Ising model, the uniform susceptibility
$\chi\sim 1/T$ for all orders of the linked cluster expansion -- we
demonstrate this explicitly.  The susceptibility in the Heisenberg
model also increases with decreasing temperature (up to the lowest
temperature we can access).  These two cases differ starkly from the
XY model where the uniform susceptibility (for magnetization in the
$z$-direction, i.e., the same quantity calculated in the other two
models) shows a plateau at low temperatures. At high temperatures, the
clean and disordered models behave identically as regards the energy,
specific heat, and entropy up to third order in inverse temperature,
and show identical susceptibilities up to second order -- we show this
explicitly in Sec.~\ref{sec:results} via a high temperature expansion.

The presentation is organized as follows.  In section
\ref{sec:models}, we introduce the three models we study (the spin-1/2
Ising, XY, and Heisenberg) and summarize some of their known
properties in the square and honeycomb lattices.
Section~\ref{sec:nlce} briefly describes NLCEs for systems with
disorder.  Numerical results for the latter, and their comparison with
those for clean systems, are presented in Sec.~\ref{sec:results}. We
conclude with a brief summary in Sec.~\ref{sec:summary}.

\section{Models}
\label{sec:models}

\begin{figure}[!b]
  \centering
  \includegraphics[width=0.95\columnwidth]{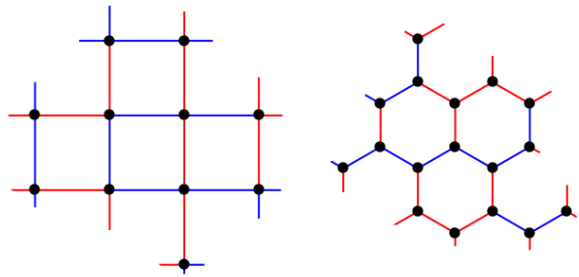}
  \caption{(Color online) Schematic of lattice models, square (left) and 
  honeycomb (right), with bond disorder considered in this work. 
  The red and blue bonds represent $J_{ij}=\pm J$. Black dots at vertices 
  represent the spins. It is easy to see that the bond disorder causes 
  frustration by trying to arrange the spins to minimize energy.}
  \label{fig:schematic}
\end{figure}

We are interested in thermodynamic properties of various spin-1/2
models on the square and honeycomb lattices (see Fig.~\ref{fig:schematic}). 
In the absence of disorder, and for nearest neighbor interactions, those 
models do not exhibit frustration on either lattice, which are both 
bipartite. While the thermodynamic properties of the various models 
studied here are qualitatively similar on both lattice geometries, 
there are significant quantitative differences, e.g., in the critical
temperatures for the onset of quasi-long-range order.\cite{sykes62}
These differences have their origin in the different coordination
number in both lattices, with the honeycomb lattice having the
smallest one. Hence, not surprisingly, for the spin-1/2
antiferromagnetic Heisenberg model, the staggered magnetization is
significantly suppressed on the honeycomb lattice as compared with the
square lattice.\cite{weihong91} Disorder, on the other hand, leads to
frustration on both lattices, and to a qualitative change of the
intermediate to low temperature properties with respect to the clean
systems. Frustration can be easily identified by trying to 
assign spins to the various sites in Fig.~\ref{fig:schematic} to minimize the
energy. If one takes the Ising Hamiltonian discussed below, one finds that 
for the overwhelming majority of disorder realizations there is no single
spin configuration that minimizes the energy on all bonds.

We should stress that both for the clean and disordered cases, we
expect quantum fluctuations to strongly modify the results for the
spin-1/2 XY and Heisenberg models from their classical counterparts
(see, e.g., Ref.~\onlinecite{diciolo_carrasquilla_14} for examples of
the effects of quantum fluctuations in frustrated spin-1/2 XY and
Heisenberg models on the honeycomb lattice). Our focus in this work 
is on systems with bimodal random-bond disorder, where the nearest-neighbor 
coupling between the spins takes values $\pm J$ with equal probability.

\subsection{Ising model}
\label{sec:ising}

The Hamiltonian for the spin-1/2 Ising model can be written as
\begin{equation}
  \label{eq:ising-ham}
  H_{\rm Ising} = \sum_{\langle i j \rangle} J_{ij}\, S^{z}_{i}S^{z}_{j}
\end{equation}
where $S^z_{i}=\pm 1/2$ is the spin at site-$i$, and the sum is over
nearest neighbors. In the absence of disorder ($J_{ij}=J$ for all
$i,j$), Eq.~\eqref{eq:ising-ham} has served as the quintessential
model for magnetism and was solved exactly on a 2D square lattice by
Onsager.\cite{Onsager_44} In the presence of continuous disorder, the
Hamiltonian in Eq.~\eqref{eq:ising-ham}, commonly known as the
Edwards-Anderson model,\cite{Edwards_75} has also become a widely
studied model for spin-glasses.

The Ising model has a discrete $\mathbb{Z}_{2}$ symmetry, i.e., the
transformation $S^{z}_{j}\to -S^{z}_{j}$ for all $j$ leaves the
Hamiltonian invariant. This symmetry can be spontaneously broken at
sufficiently low temperatures to create an ordered phase. For
$S^{z}_{i}=\pm 1/2$, the critical temperature is given by $T_{c}/|J| =
1/2\log(1+\sqrt{2}) \approx 0.57$ for the square lattice, and
$T_{c}/|J| = 1/2\log[(\sqrt{3}+1)/(\sqrt{3}-1)] \approx 0.38$ for the
honeycomb lattice, and is the same for the ferro- and
antiferromagnetic cases.\cite{sykes62} The latter is because, for
bipartite lattices, a unitary transformation relates both models. This
can be easily seen by rewriting the Hamiltonian in
Eq.~\eqref{eq:ising-ham} as $H_{\rm Ising} = \sum_{i}J_{i}\,
S^{z}_{i,A}S^{z}_{i,B}$, where $A$ and $B$ are the two sub-lattice
indices.  One can then go from $J_{i} \to -J_{i}$, i.e., between the
ferro- and antiferromagnetic models, via the transformation:
$S^{z}_{i,A} \to -S^{z}_{i,A}$ and $S^{z}_{i,B} \to S^{z}_{i,B}$.

The specific heat of the clean model diverges at the critical
temperature for both the square and the honeycomb
lattices. Equivalently, the derivative of the energy diverges at the
critical temperature but the energy remains finite throughout.  For
the antiferromagnetic model, the susceptibility is known to be finite
everywhere, with an infinite slope at the critical temperature. The
maximum value of the susceptibility occurs at $T_{m}=1.537\,T_{c}$ and
$T_{m}=1.688\,T_{c}$ for the square and honeycomb models
respectively,\cite{fisher59,sykes62} i.e., above the critical
temperature. Our results for the clean systems are consistent with
these. However, we cannot study the critical phase or the properties
of the system very close to criticality (see Figs.~\ref{fig:ising-sq} and
\ref{fig:ising-hc}).

The Ising model with bimodal disorder has been extensively studied in
the past.\cite{cho97,Blackman1998,hartman01,gruzberg01,Amoruso_04,
  Lukic04,Katzgraber05,Katzgraber07,atisattapong09,thomas_huse_11,parisen_francesco_11}
It is reasonably well established that no glassy phase exists for
$T>0$.\cite{hartman01,gruzberg01} It has, however, been established that
a glassy phase appears at zero temperature in this model.\cite{thomas_huse_11}  At finite temperature, our results
for this case are described in Sec.~\ref{sec:ising-results}.

\subsection{XY Model}
\label{sec:xy}

The spin-1/2 XY model can be written as
\begin{equation}
  \label{eq:XY}
  \hat{H}_{\rm XY} = \sum_{\langle i j \rangle} J_{ij} 
  (\hat{S}^{x}_{i}\hat{S}^{x}_{j} + \hat{S}^{y}_{i}\hat{S}^{y}_{j})
\end{equation}
where $\hat{S}^{x,y}_{i}$ are spin operators at site $i$, proportional
to the Pauli matrices.  We consider only the isotropic case, where the
model has a continuous $U(1)$ rotation symmetry in the plane. The
presence of a continuous symmetry precludes, via the Mermin-Wagner
theorem,\cite{mermin66} a finite-temperature phase-transition
involving the breaking of this continuous symmetry from occurring. For
$d\leq 2$ dimensions, the fluctuations in any putative ordered phase
appearing from breaking a continuous symmetry grow with system size
for finite temperatures, destroying any order (see e.g.,
Ref.~\onlinecite{altland06}).  Hence, this model has an ordered phase
only at $T=0$.

However, in 2D there can still be a finite temperature
Berezinskii-Kosterlitz-Thouless (BKT)
transition\cite{berezinski72,kosterlitz73} below which the system
exhibits quasi-long-range spin order. The critical temperature for the
BKT transition in the spin-1/2 XY model in the square lattice is
$T_{c}/|J|\approx 0.34$.\cite{harada98,carrasquilla12} We should
stress that both classical\cite{berezinskii71} and quantum models with
continuous symmetries in 2D can exhibit this kind of
behavior.\cite{frohlich81} We should also mention that most studies in
the literature report results for the ferromagnetic XY model
($J<1$). However, like for the Ising model, in the square and
honeycomb lattice a unitary transformation relates the ferro- and
antiferromagnetic models and the critical temperature is the same in
both. Our calculations for the susceptibility of the clean case on the
square lattice converge down to temperatures of $T/|J|\approx 0.4$,
which is compatible with the onset of quasi-long-range order for
$T_{c}/|J|\lesssim 0.34$.\cite{harada98,carrasquilla12}

\subsection{Heisenberg model}
\label{sec:heisenberg}

The spin-1/2 Heisenberg model, also known as the XXX model, can be
written as
\begin{equation}
  \label{eq:XXX}
  \hat{H}_{\rm Heis} = \sum_{\langle i j \rangle} J_{ij}\, 
  \hat{\mathbf{S}}_{i}\cdot \hat{\mathbf{S}}_{j}
\end{equation}
where $\hat{\mathbf{S}}_i =
(\hat{S}^{x}_i,\hat{S}^{y}_i,\hat{S}^{z}_i)$, and
$\hat{S}^{x,y,z}_{i}$ are spin operators at site $i$, proportional to
the Pauli matrices. The Heisenberg model has an $SU(2)$ symmetry, the
highest of the three models considered in this work. The ground state
in the clean case ($J_{ij}=J$) is an ordered ferromagnet or
antiferromagnet depending on the sign of the coupling constant
$J$. Like for the XY model, long-range order only occurs at zero
temperature.  However, in contrast to the XY model, the 2D Heisenberg
model does not develop quasi-long-range order at finite
temperature. This is due to the fact that the internal symmetry group,
$SU(2)$ [$O(3)$] for the quantum (classical) Heisenberg model, is
non-Abelian, as opposed to the XY model which has an Abelian symmetry
group, $U(1)$ [$O(2)$] for the quantum (classical) cases. Vortices or
point-defects, which are responsible for the BKT
transition,\cite{kosterlitz73} occur only in the latter
case.\cite{kenna05} Furthermore, in 2D, $O(N)$ models with $N\geq 3$
or $SU(N)$ models with $N\geq 2$, i.e., with non-Abelian symmetry
groups, can be shown via perturbation theory to be asymptotically
free, which for spin models translates to a renormalization group
flow towards paramagnetism.\cite{gross73,gross99}

The results for the clean system presented here for the square lattice
are nearly identical to the spin-1/2 results presented in
Ref.~\onlinecite{cuccoli97}, which also compares with other known
results.

\section{Numerical Linked Cluster Expansions}
\label{sec:nlce}

Numerical linked cluster expansions (NLCEs) are a computational
technique that can be used to calculate extensive properties (per
lattice site) of translationally invariant lattice systems. NLCEs, which are based
on linked cluster expansions,\cite{domb60a,domb60b,sykes66} were
introduced in Ref.~\onlinecite{rigol_bryant_06}, where it was shown
that the results obtained for thermodynamic properties were exact in
the thermodynamic limit for systems with a finite correlation
length. Furthermore, results could be obtained at significantly lower
temperatures as compared to high-temperature expansions for models
that develop long-range order at zero temperature.  In several
subsequent works, NLCEs have been shown to be a powerful computational
technique for determining not only thermodynamic properties of a
variety of lattice models,\cite{rigol_bryant_07,rigol_bryant_07b,
  khatami11a,khatami11b,tang12,tang13b} but also for studying
thermalization (or the lack thereof) at long times after a quench in
isolated quantum
systems.\cite{rigol_14a,wouters_denardis_14,rigol_14b,tang2014} For
completeness, we provide a brief description of NLCEs. Details of how
to implement them can be found in Ref.~\onlinecite{tang_khatami_13}.

In NLCEs, the expectation value of an extensive observable,
\emph{per-site}, $\mathcal{O}$ in a translationally invariant system
can be calculated as a sum over contributions from all clusters $c$ of
different sizes that can be embedded on the lattice
\begin{equation}
  \label{eq:nlce}
  \mathcal{O} = \sum_{c}M(c)\times W_{\mathcal{O}}(c),
\end{equation}
where $M(c)$ is a combinatorial factor equal to the number of ways
that a particular cluster $c$ can be embedded on that
lattice. $W_{\mathcal{O}}(c)$ is the \emph{weight} of cluster $c$ for
the given observable, which is calculated via the inclusion-exclusion
principle
\begin{equation}
  W_{\cal O}(c) =
  \mathcal{O}(c) - \sum_{s\subset c}W_{\cal O}(s).
\end{equation} 
$\mathcal{O}(c)$ is the expectation value of the observable
$\mathcal{O}$ on the specific cluster $c$. Within NLCEs,
$\mathcal{O}(c)$ is calculated using full exact diagonalization. The
expansion is carried out order by order, i.e., by first considering
clusters with one-site, then two-sites, and so on.  Beyond the
\emph{bare} sum in Eq.~\eqref{eq:nlce}, several resummation schemes
exist that accelerate the convergence of the
expansion.\cite{rigol_bryant_07} Here we will report results from Wynn
and Euler resummation techniques whenever they allow us to extend the
convergence of the results to lower
temperatures.\cite{rigol_bryant_07}

\begin{figure}[!t]
  \centering
  \includegraphics[width=0.6\columnwidth]{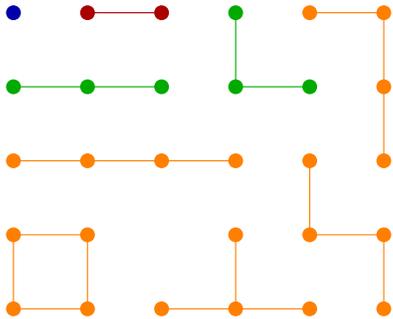}
  \caption{(Color online) Clusters up to the fourth order in the site
    based NLCE on the square lattice. The two 3-site clusters have the
    same Hamiltonian. At fourth order, in addition to three clusters
    with identical Hamiltonian, two topologically new clusters appear
    -- the closed loop and the ``$\perp$''. Each topologically new
    cluster is diagonalized separately.}
  \label{fig:clusters}
\end{figure}

Examples of clusters up to fourth order in the site-based NLCE used
here on the square lattice are shown in Fig.~\ref{fig:clusters}.  At
third order, although there are two \emph{geometrically} different
clusters, they are \emph{topologically} identical. They have the same
Hamiltonian for the models with nearest interactions we consider
here. At fourth order, there are three clusters (including the one
with the four sites on a line) that again have the same Hamiltonian
for the models considered here. However, two \emph{topologically} new
clusters appear, namely, the closed loop and the ``$\perp$''. They
have to be individually diagonalized.  From the fourth order and
beyond the number of distinct topological clusters increases rapidly
(exponentially with the number of sites), making the calculations
increasingly costly. References~\onlinecite{rigol_bryant_07,tang13b}
provide details on the various topological clusters on the square
and honeycomb lattices, respectively, as well as the number of such
clusters as a function of the order of the expansion.

Recently, in Ref.~\onlinecite{tang2014}, it was shown that NLCEs can
also be used to study systems with disorder. As described above, NLCEs
can only be used for translationally invariant systems. \emph{A
  priori}, disorder breaks translational invariance. However, we are
only interested in disorder averaged physical quantities. If we take a
disorder average over all possible disorder configurations in models
with bimodal disorder, we restore translational invariance and the
equivalent of Eq.~\eqref{eq:nlce} reads
\begin{equation}
  \label{eq:nlce-dis}
  \overline{\mathcal{O}} = \sum_{c}M(c)\times \overline{W_{\mathcal{O}}(c)}
\end{equation}
where the overline denotes the disorder average. The disorder average
of the weights is in turn given by
\begin{equation}
  \overline{W_{\cal O}(c)} = \overline{\mathcal{O}(c)} 
  - \sum_{s\subset c}\overline{W_{\cal O}(s)}.
\end{equation}
In other words, the disorder average can be carried out order by order
for each observable. The calculations then proceed as for the
translationally invariant system if one replaces $\mathcal{O}(c)$ by
$\overline{\mathcal{O}(c)}$.

The computations in the presence of disorder are much more challenging
than for translationally invariant systems because of the additional
average over all possible disorder realizations. For example, the
largest clusters we consider here for the quantum models in the square
lattice have 13 sites.  At this order, there are a total of $\sim
1.9\times 10^{6}$ connected clusters, of which 5,450 are topologically
distinct.\cite{rigol_bryant_07} Each of these has to be fully
diagonalized for the $2^{\ell}$ disorder configurations corresponding
to the $\ell$ bonds in the cluster. This has to, of course, be carried
out for all lower orders as well, each with a different set of
topologically distinct clusters and disorder configurations.  For the
clean systems, we report results for cluster with up to 15 sites for
the square lattice. In that case one has to diagonalize 42,192
topologically distinct clusters with 15 sites.

NLCE calculations fail to converge when correlations in the
thermodynamic limit extend beyond the largest clusters
considered. Therefore, NLCEs cannot be used to calculate observables
in phases with long-range order unless one tailors the expansion to
account for those.\cite{khatami_singh_11} Since disorder usually
shortens correlations at low temperatures, NLCEs are particularly
useful to study quantum disordered systems, despite the increase of
computational cost because of the disorder average. It will become
apparent when we discuss our results for the various thermodynamic
properties of interest in this work, that NLCEs converge to lower
temperatures in disordered systems when compared to clean systems. As
mentioned before, quantum Monte Carlo simulations in the presence of
disorder are severely limited by the sign problem.

\section{Results}
\label{sec:results}

In this section, we discuss the results of our NLCE based study of the
three spin models described in Sec.~\ref{sec:models} on the square and
honeycomb lattices.  In what follows, we set $J=1$ as the energy
scale. For each model and lattice geometry, we report the energy
($E$), entropy ($S$), specific heat ($C_{v}$), and uniform
susceptibility for the magnetization along the $z$-axis ($\chi$) as a
function of temperature.  These quantities are defined as
\begin{equation}
  \begin{split}
    \label{eq:definitions}
    E = \frac{\overline{\langle \hat{H}\rangle}}{N},\qquad &C_{v} =
    \frac{\overline{\langle \hat{H}^{2}\rangle} -
      \overline{\langle \hat{H}\rangle^{2}}}{NT^{2}} \\
    S = \frac{\overline{\log Z}}{N} + \frac{E}{T}, \qquad &\chi =
    \frac{\overline{\langle (\hat{S}^{z})^{2}\rangle} -
      \overline{\langle \hat{S}^{z}\rangle^{2}}}{NT}
  \end{split}
\end{equation}
where the overline denotes a disorder average, the angle brackets
denote the thermal expectation value in the grand-canonical ensemble
(at zero chemical potential), $N$ is the number of sites, and $T$ is
the temperature. As mentioned earlier, the disorder average is carried
out over all disorder configurations at each order in the NLCE.  In
all cases, the disorder averaged results are compared with the ones in
clean systems.

For all observables, we report bare NLCE results for the highest two
orders of our site-based NLCE expansion, which are determined by the
number of sites $l$ in the largest clusters studied. Namely, we report
the results from Eq.~\eqref{eq:nlce-dis} when the contributions of all
clusters with up to $l$ sites are added, where $l$ takes the two
largest values in our calculations for each model in each lattice
geometry.  We also report results using two different resummation
schemes, indicated as Wynn$_{n}$ and Euler$_{n}$. The subscript $n$
denotes the order of the resummation process (see
Ref.~\onlinecite{rigol_bryant_07} for details).  The resummation
schemes allow us to access significantly lower temperatures than the
bare results in some cases as indicated below.

\begin{figure}[!b]
  \includegraphics[width=1\columnwidth]{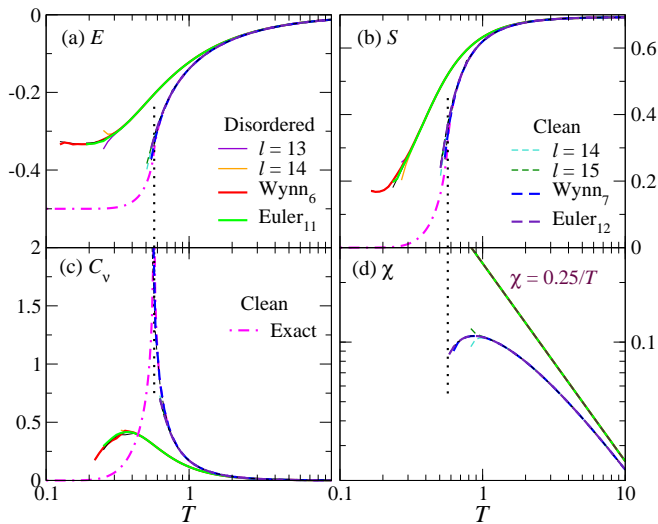}
  \caption{(Color online) \textbf{Spin-1/2 Ising model on the square
      lattice}. Panels (a)--(d) show the energy, entropy, specific
    heat, and uniform susceptibility vs $T$, respectively.  Solid
    lines depict disorder averaged quantities, while dashed lines
    depict results for the clean system. Thin lines report bare
    results for the last two orders of the NLCE, while thick lines
    report the results of two resummation techniques. A thin
    continuous line following the results of the resummations reports
    results for a lower order of the same resummation technique and is
    used to gauge their stability. The dotted vertical line marks the
    position of the phase transition. The dashed dotted line shows
    exact analytic results for the clean system in the thermodynamic
    limit.}
  \label{fig:ising-sq}
\end{figure}

\begin{figure}[!b]
  \centering
  \includegraphics[width=1\columnwidth]{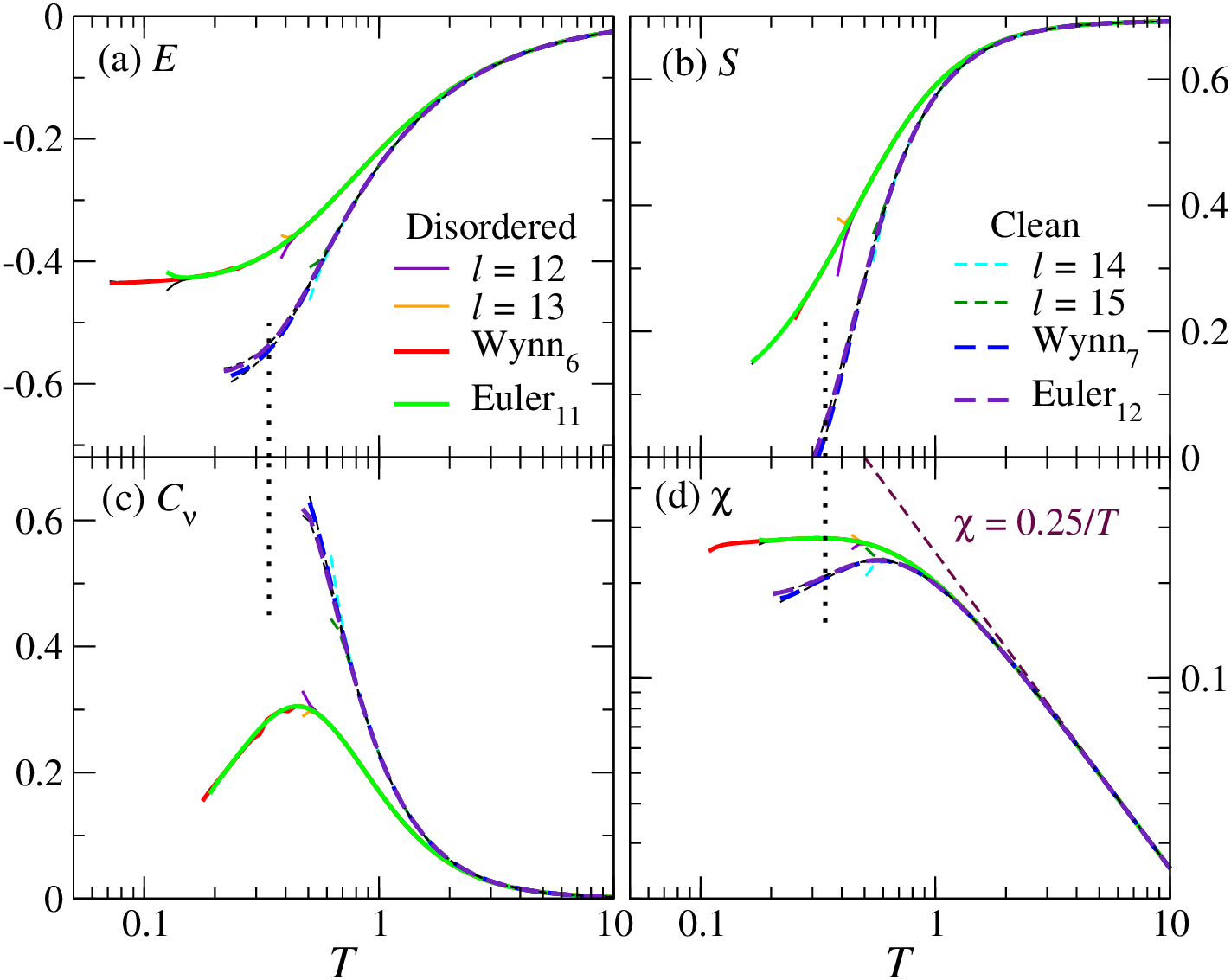}
  \caption{(Color online) \textbf{Spin-1/2 XY model on the square
      lattice}. Panels (a)--(d) show the energy, entropy, specific
    heat, and uniform susceptibility vs $T$, respectively.  Solid
    lines depict disorder averaged quantities, while dashed lines
    depict results for the clean system. Thin lines report bare
    results for the last two orders of the NLCE, while thick lines
    report the results of two resummation techniques. A thin
    continuous line following the results of the resummations reports
    results for a lower order of the same resummation technique and is
    used to gauge their stability. The dotted vertical line marks the
    position of the phase transition.}
  \label{fig:xy-sq}
\end{figure}

\begin{figure}[!b]
  \centering
  \includegraphics[width=1\columnwidth]{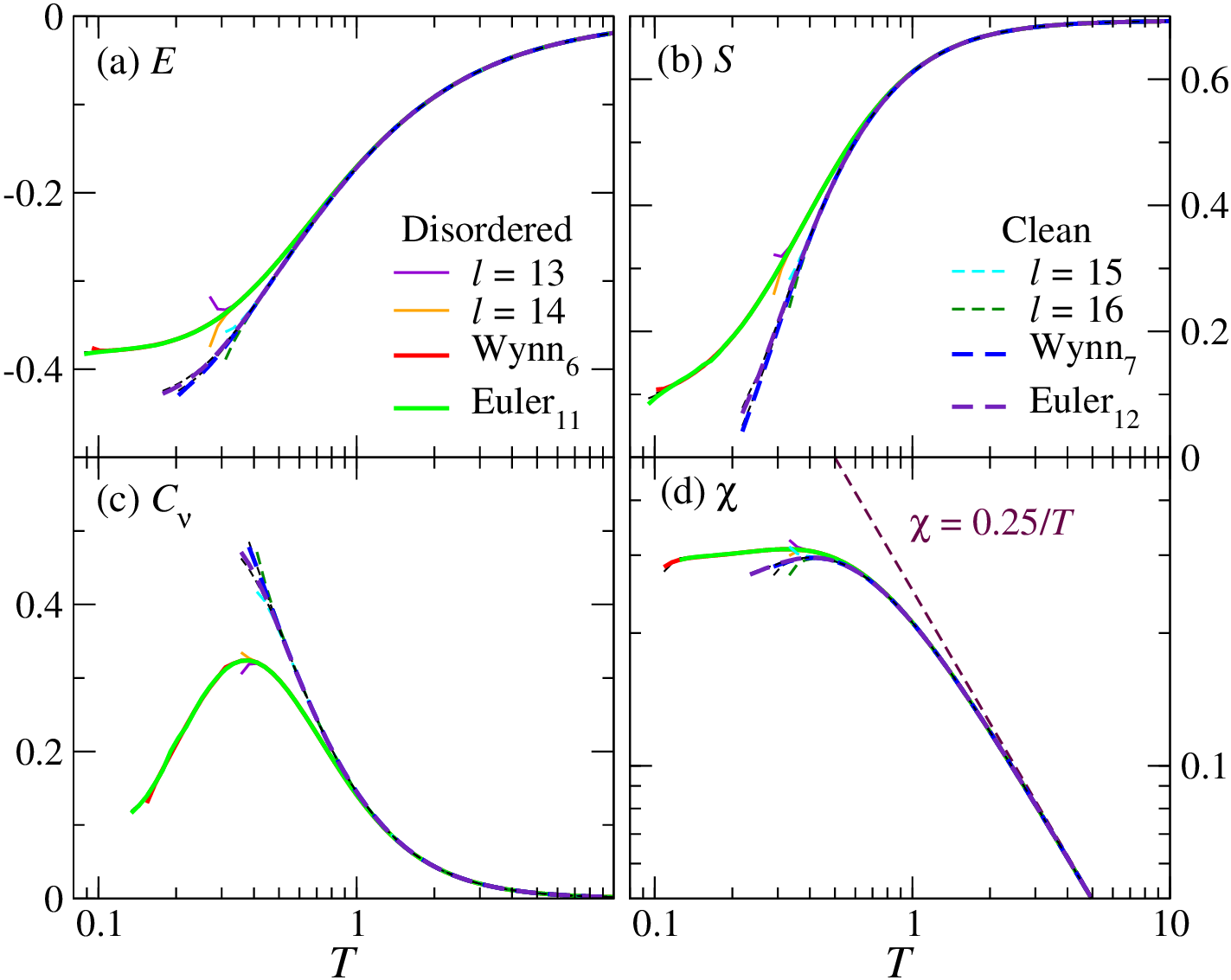}
  \caption{(Color online) \textbf{Spin-1/2 XY model on the honeycomb
      lattice}. Panels (a)--(d) show the energy, entropy, specific
    heat, and uniform susceptibility vs $T$, respectively.  Solid
    lines depict disorder averaged quantities, while dashed lines
    depict results for the clean system. Thin lines report bare
    results for the last two orders of the NLCE, while thick lines
    report the results of two resummation techniques. A thin
    continuous line following the results of the resummations reports
    results for a lower order of the same resummation technique and is
    used to gauge their stability. Note that the results converge to
    similar temperatures as in the square lattice.}
  \label{fig:xy-hc}
\end{figure}

Before discussing each model and observable in detail, we review a few
general observations for completeness and pedagogy. In all models and
observables discussed here, the numerical results at intermediate to
high temperatures in the presence of disorder are close to those of
the corresponding clean system.  Whereas this is obvious for
temperatures so high that the first order correction to the infinite
temperature result is negligibly small, we notice from our results in
Figs.~\ref{fig:ising-sq}--\ref{fig:xxx-hc} that the observables in
clean and disordered models are indistinguishable for temperatures as
low as $T=2$ (barring the susceptibility).

In order to show why this is so, we expand the partition function for
small inverse temperature $\beta\equiv 1/T$,
$Z=\Tr(e^{-\beta\hat{H}})\approx\Tr(1-\beta\hat{H}+\beta^{2}\hat{H}^{2}/2+\ldots)$.
The models we consider have only nearest-neighbor coupling, i.e.,
$\hat{H}=\sum_{\langle ij\rangle}[J_{ij}\hat{H}_{ij}+h(\hat{S}^{z}_{i}
+\hat{S}^{z}_{j})/2]$, where we have included a magnetic field $h$ as
a source to calculate the uniform susceptibility in the
$z$-direction. The most general two-site Hamiltonian that describes
all models of interest here is given by $H_{ij} =
\gamma(\hat{S}^{x}_{i}\hat{S}^{x}_{j} +
\hat{S}^{y}_{i}\hat{S}^{y}_{j}) + \Delta
\hat{S}^{z}_{i}\hat{S}^{z}_{j}$, which becomes the Ising model for
$\gamma=0,\Delta=1$, the XY model for $\gamma=1,\Delta=0$, and the
Heisenberg model for $\gamma=\Delta=1$.  With this in mind, to first
order in $\beta$, the high temperature expansion for $Z$ can be
written as $Z = 2^{N} - \beta\sum_{\langle i j
  \rangle}\Tr[J_{ij}\hat{H}_{ij} +
h(\hat{S}^{z}_{i}+\hat{S}^{z}_{j})/2]$, where $N$ is the number of
lattice sites.  Note first that $\Tr(\hat{S}^{z}_{i})=0$ (the Pauli
matrices are traceless), and second, that $\Tr(\hat{H}_{ij})=0$, so
that the linear correction vanishes. Therefore, to first order in
$\beta$, the clean and the disordered system have the same partition
function.  This is true regardless of the type of disorder.

To second order, after expanding $\hat{H}^{2}$, we have
\begin{multline}
  \label{eq:hte-2-z}
  Z = 2^{N} + \frac{\beta^{2}}{2} \Tr \Bigg[\sum_{\langle i j
    \rangle,\langle k l \rangle} J_{ij}J_{kl}\hat{H}_{ij}\hat{H}_{kl}+
  \\ + h^{2}\sum_{i,j}\hat{S}^{z}_{i}\hat{S}^{z}_{j} + h\sum_{\langle
    i j \rangle,k} J_{ij}\hat{H}_{ij}\hat{S}^{z}_{k}\Bigg]
\end{multline}
Let us treat the above terms one by one.  In the first term, for
$\langle i j \rangle \neq \langle k l \rangle$, the trace is
identically zero as shown above.  For the case when say $i\neq l$, but
$j=k$, we effectively have a new Hamiltonian for three neighboring
spins. It is easy to verify explicitly that this trace also
vanishes. The only possibility left for a nonvanishing contribution is
$\langle i j \rangle = \langle k l \rangle$.  The trace of second term
in the brackets is nonzero only for $i=j$.  In the third term, again
for $k\neq i\text{ or } j$, the trace is zero. For, say $j=k$,
considering only the diagonal elements of the matrix,
$\hat{S}^{z}_{i}\hat{S}^{z}_{j}\hat{S}^{z}_{j} \propto
\hat{S}^{z}_{i}$, we see that the trace vanishes. Taking these into
account, and writing $\log Z$ to $O(\beta^{2})$, we have
\begin{equation}
  \label{eq:hte-2-logz}
  \frac{\log Z}{N} = \log 2 + \frac{\beta^{2}}{2\cdot 2^{N}N}
  \Tr \Bigg[\sum_{\langle i j \rangle} J_{ij}^{2}\hat{H}_{ij}^{2} 
  + \frac{N h^{2}}{4} \Bigg]
\end{equation}
For both, the clean system and the system with bimodal disorder,
$J_{ij}^{2}=J^{2}$. We therefore get,
\begin{equation}
  \label{eq:hte-2-logz-2}
  \frac{\log Z}{N} = \log 2 + \frac{\beta^{2}}{4}\bigg(J^{2}\Tr_{2}\hat{H}_{2}^{2} 
  + \frac{h^{2}}{2} \bigg) + O(\beta^{3})
\end{equation}
where $\hat{H}_{2}$ is the Hamiltonian for a two-site system and the
subscript ``2'' on the trace indicates a trace over the Fock space of
a two-site system.  The disorder average does not change the above
expression, and therefore to this order, the clean and the disordered
systems behave identically.  The energy, for instance, is given to
this order by $E=-(\partial\log Z/\partial\beta)/N = -\beta
J^{2}\Tr_{2}(\hat{H}_{2}^{2})/2$, the specific heat is
$C_{v}=\beta^{2}J^{2}\Tr_{2}(\hat{H}_{2}^{2})/2$, and the uniform
susceptibility is $\chi = \beta/4$.

In fact, it is straightforward to check that the partition function in
the clean and disordered systems remains the same at
$O(\beta^{3})$, except for terms proportional to $h^{2}$. In other
words, the energy, entropy, and specific heat are the same for clean
and disordered systems up to third order in $\beta$, but the uniform
susceptibility deviates. The following expression can be derived along
the lines of Eqs.~\eqref{eq:hte-2-z}--\eqref{eq:hte-2-logz-2} for the
third order correction to the partition function:
\begin{multline}
  \label{eq:hte-3-logz}
  \frac{\log Z}{N} = \log 2 +
  \frac{\beta^{2}}{4}\bigg(J^{2}\Tr_{2}\hat{H}_{2}^{2}+
  \frac{h^{2}}{2} \bigg) - \\ - \frac{\beta^{3}h^{2}}{12}
  J_{ij}\Tr_{2} \hat{H}_{ij}\hat{S}^{z}_{i}\hat{S}^{z}_{j}.
\end{multline}
There is no sum over $i,j$, which represent the two sites in a 2-site
system.  For the clean model one just needs to replace $J_{ij}$ with
$J$ in the above formula, while for the disordered one, the disorder
average produces two terms for $\pm J$ (which cancel each other,
implying that disorder extends the paramagnetic behavior in the
susceptibility to lower temperatures).  One can the see that for
$h=0$, all thermodynamic quantities studied here, except the
susceptibility, are identical in the clean and disordered cases up to
$O(\beta^{3})$. One can further verify that this changes at fourth
order, where differences emerge between clean and disordered
systems. An example of a term that makes a difference at fourth order
is a square loop with four sites and four bonds (see
Fig.~\ref{fig:clusters}). Even in the Ising case, the Hamiltonian
$H_{12}H_{23}H_{34}H_{41}=1/64$ has a nonzero trace with a product of
four different $J_{ij}$.

We should stress that, in all models studied here in the presence of
disorder, we find that there is a significant amount of residual
entropy (when comparing with the clean systems) at the lowest
temperatures we are able to access with NLCEs. This is a clear
indication of the lack of order at those temperatures. The behavior of
the entropy, coupled with a saturation of the energy observed at the
lowest temperatures accessible to us, confirms that there are many
energy levels close to each other at low energies. This is the
hallmark of frustration.

\subsection{Ising model}
\label{sec:ising-results}

\begin{figure}[!b]
  \includegraphics[width=1\columnwidth]{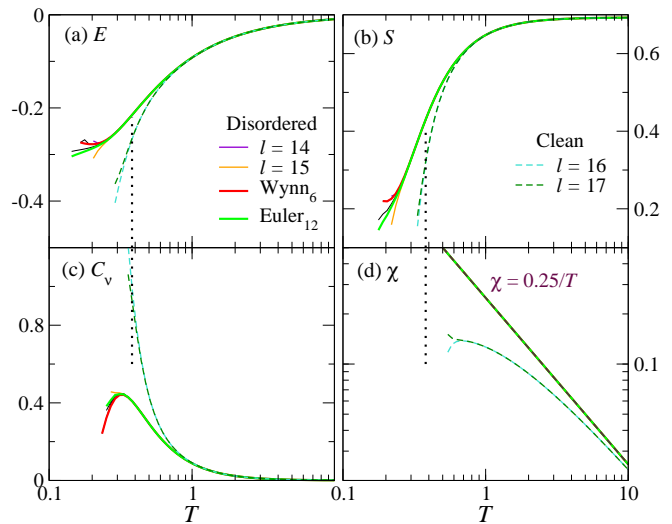}
  \caption{(Color online) \textbf{Spin-1/2 Ising model on the
      honeycomb lattice}. Panels (a)--(d) show the energy, entropy,
    specific heat, and uniform susceptibility vs $T$, respectively.
    Solid lines depict disorder averaged quantities, while dashed
    lines depict results for the clean system. Thin lines report bare
    results for the last two orders of the NLCE, while thick lines
    report the results of two resummation techniques. A thin
    continuous line following the results of the resummations reports
    results for a lower order of the same resummation technique and is
    used to gauge their stability. Resummation results are not
    presented for the clean case as they do not extend the convergence
    to lower temperatures. The dotted vertical line marks the position
    of the phase transition.}
  \label{fig:ising-hc}
\end{figure}

Figures \ref{fig:ising-sq} and \ref{fig:ising-hc} show the energy (a),
entropy (b), specific heat (c), and uniform susceptibility (d) for the
disordered spin-1/2 Ising model on the square and honeycomb lattices,
respectively. For the square lattice, we also plot the exact results
for $E$, $S$, and $C_{v}$ for the clean
model.\cite{Onsager_44,huang87} Several approximate analytic estimates
exist for the susceptibility, but there is no closed form expression
for all temperatures.

Figures \ref{fig:ising-sq}(a) and \ref{fig:ising-hc}(a) show that, as
mentioned before, a generic feature in the presence of disorder is
that the energy tends to plateau at the lowest temperatures accessible
to us. In that regime, the entropy is significantly higher than in the
clean systems [Figs.~\ref{fig:ising-sq}(b) and
\ref{fig:ising-hc}(b)]. Distinct to the Ising models, the sharp
divergence in the specific heat in the clean case [see
Figs.~\ref{fig:ising-sq}(c) and \ref{fig:ising-hc}(c)], which
indicates the phase transition, is replaced by what appears to be
smooth peak in the presence of disorder. The maximum of that peak
appears at temperatures lower than the critical temperature in the
clean case. At higher temperatures, Eq.~\eqref{eq:hte-2-logz-2} gives
results that agree for $E$, $C_{v}$, and $S$ down to $T\approx 1$. We
note that NLCE results for the disordered model are well converged
down to $T\approx 0.2$ to 0.3, while the NLCE results for the clean
model converge to temperatures that are close to $T_c$, and agree with
the analytic results in the disordered phase.

Results for the uniform susceptibility in Figs.~\ref{fig:ising-sq}(d)
and \ref{fig:ising-hc}(d) show that this quantity behaves very
differently in the clean and disordered systems. In the disordered
case it exhibits a $1/T$ behavior at all temperatures, both on the
square and honeycomb lattices.  An order by order linked cluster
expansion reveals that the only nonvanishing contribution to the
susceptibility in the disordered case comes from the single-site
system, and is trivially proportional to $1/T$.  All higher order
contributions vanish. We show this for the first few orders of the
NLCE on the square lattice.  Below are the expressions for the
disorder averaged $\log Z$ for the clusters with one, two, and three
sites shown in Fig.~\ref{fig:clusters}.

\newpage

\begin{widetext}
  \begin{equation}
    \label{eq:ising-logz}
    \begin{split}
      \log Z^{(1)} &= \log\left(e^{-\frac{\beta h}{2}}+e^{\frac{\beta h}{2}}\right) \\
      \log Z^{(2)} &= \frac12\left[ \log\left(2e^{-\frac{\beta J}{4}}
          + e^{\frac{\beta(J+4h)}{4}} + e^{\frac{\beta(J-4h)}{4}}
        \right)
        +    \beta\to -\beta \right]\\
      \log Z^{(3)} &= \frac12\log\left(e^{\frac{\beta h}{2}} +
        e^{\frac{3\beta h}{2}} +
        e^{\frac{\beta(J+h)}{2}} + e^{\frac{\beta(J-h)}{2}} +  \beta\to -\beta \right) + \\
      &\qquad + \frac14\left[ \log\left(2e^{\frac{\beta h}{2}} + 2
          e^{\frac{-\beta h}{2}} + e^{\frac{-\beta(J+3h)}{2}} +
          e^{\frac{-\beta(J-3h)}{2}} + e^{\frac{\beta(J+h)}{2}} +
          e^{\frac{\beta(J-h)}{2}} \right) + \beta \to -\beta \right]
    \end{split}
  \end{equation}
\end{widetext}

The uniform susceptibility can be obtained from $\chi =
\beta^{-1}(\partial^{2}\log Z)/(\partial h)^{2}$ evaluated at $h=0$.
We get for the above three orders,
\begin{equation}
  \label{eq:ising-chi}
  \chi^{(1)} = \frac{\beta}{4}, \quad
  \chi^{(2)} = \frac{2\beta}{4}, \quad
  \chi^{(3)} = \frac{3\beta}{4}.
\end{equation}
Already, one can see that no new contributions appear at the higher
orders. To confirm this, we calculate the weights of the three
clusters (see Ref.~\onlinecite{rigol_bryant_07} for details about
multiplicities, etc.):
\begin{equation}
  \label{eq:ising-weights}
  \begin{split}
    W_{\chi}^{(1)} &= \chi^{(1)} = \frac{\beta}{4} \\
    W_{\chi}^{(2)} &= \chi^{(2)}-2W_{\chi}^{(1)} = 0 \\
    W_{\chi}^{(3)} &= \chi^{(3)} - 2W_{\chi}^{(2)} - 3W_{\chi}^{(1)} =
    0.
  \end{split}
\end{equation}
Indeed, only the single-site cluster contributes. One can check this
at higher orders, and for the honeycomb lattice as well. This is not
the case for the XY and Heisenberg models discussed below.

Earlier studies for the Ising model with bimodal disorder on the
square lattice found a low-temperature scaling of the specific heat
(i.e., the exponent $\alpha$ in a power law fit of the low-temperature
specific heat, $C_{v}\sim T^{-\alpha}$) that is different from the
model with continuous disorder.\cite{Katzgraber07} At even lower
temperatures, a crossover in the scaling behavior of $C_{v}$ has been
reported.\cite{thomas_huse_11} Unfortunately, our results do not
converge at low enough temperatures to observe such power
laws. However, for $T>0.3$, our results are consistent with those in
other studies\cite{Katzgraber07,thomas_huse_11} (since we consider
$S^{z}=\pm 1/2$, our temperatures are lower by a factor of four from
those studies, which took $S^{z}=\pm 1$).

\subsection{XY model}
\label{sec:xy-results}

\begin{figure}[!b]
  \centering
  \includegraphics[width=1\columnwidth]{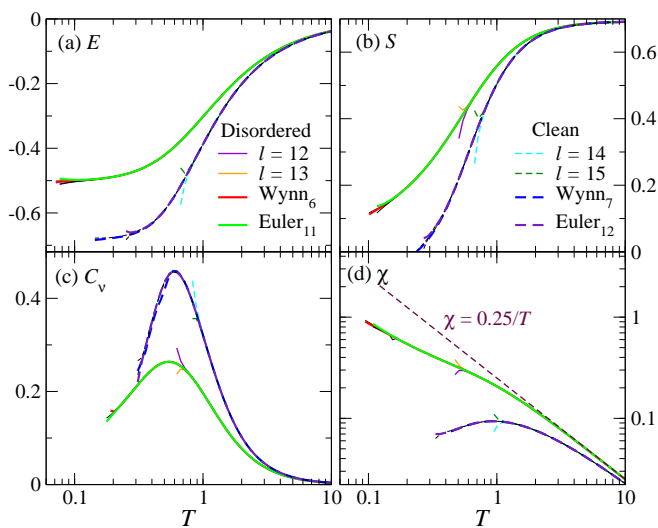}
  \caption{(Color online) \textbf{Spin-1/2 Heisenberg model on the
      square lattice}. Same as Fig.~\ref{fig:xy-sq} but for the
    spin-1/2 Heisenberg model.}
  \label{fig:xxx-sq}
\end{figure}

Figures~\ref{fig:xy-sq} and \ref{fig:xy-hc} show results for the
spin-1/2 XY model on the square and honeycomb lattice,
respectively. The results for all quantities are well converged down
to about $T\approx 0.1$ to 0.2. Figures \ref{fig:xy-sq}(a),(b) and
\ref{fig:xy-hc}(a),(b) show that the behavior of energy and the
entropy is qualitatively similar to the one observed in the Ising
model. However, the results for the XY model in the presence of
disorder converge at lower temperature than those for the Ising
model. For the XY model, the specific heat in the presence of disorder
exhibits a peak that is well resolved by our NLCE
[Figs.~\ref{fig:xy-sq}(c) and \ref{fig:xy-hc}(c)]. We note that the
energy, entropy, and specific heat follow the second order result in
Eq.~\eqref{eq:hte-2-logz-2} for $T>2$ in the square lattice and
$T>0.7$ in the honeycomb lattice.

Interestingly, Figs.~\ref{fig:xy-sq}(d) and \ref{fig:xy-hc}(d) show
that in the XY model the uniform ($z$-)susceptibility in the presence
of disorder exhibits a plateau for low temperatures. This is
qualitatively different from the behavior observed for the Ising
model. The fact that the response to an external magnetic field in the
$z$-direction is independent of temperature for low temperatures shows
that increasing temperature does not increase the disorder in the spin
correlations in the $z$-direction.

\begin{figure}[!b]
  \centering
  \includegraphics[width=1\columnwidth]{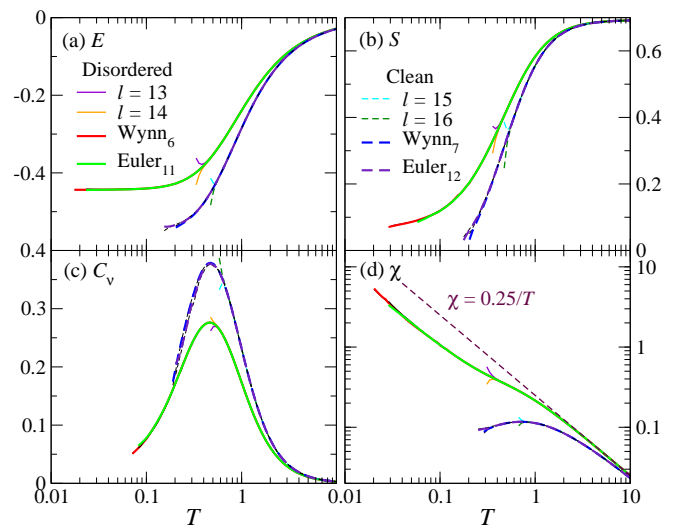}
  \caption{(Color online) \textbf{Spin-1/2 Heisenberg model on the
      honeycomb lattice}. Same as Fig.~\ref{fig:xy-hc} but for the
    spin-1/2 Heisenberg model.}
  \label{fig:xxx-hc}
\end{figure}

We should add that the classical XY model has been studied in the
presence of Gaussian-random dilution\cite{deng14} and bimodal
dilution\cite{surungan05} of ferromagnetic bonds. In these works, the
BKT transition was seen to slowly disappear as the dilution was
increased. Here we only have considered the fully disordered case,
i.e., an equal distribution of ferro- and antiferromagnetic bonds, so
we do not expect that any remnants of the BKT phase are present in our
calculations.

\subsection{Heisenberg model}
\label{sec:heisenberg-results}

Figures~\ref{fig:xxx-sq} and \ref{fig:xxx-hc} show results for the
spin-1/2 Heisenberg model on the square and honeycomb lattices,
respectively. In Figs.~\ref{fig:xxx-sq}(a) and \ref{fig:xxx-hc}(a),
one can see that the plateau in the energy at low temperatures is the
clearest of all models studied in this work. The onset of this plateau
occurs at $T\approx 0.2$ for both models. Remarkably, in the honeycomb
geometry, the energy in the presence of disorder converges down to
$T\approx0.02$. The entropy [Figs.~\ref{fig:xxx-sq}(b) and
\ref{fig:xxx-hc}(b)] behaves similarly to the entropy in the XY model,
but also converges to very low temperatures ($T\approx0.03$ to 0.04)
in the honeycomb geometry. Like for the XY model, the specific heat
exhibits a clear peak in the presence of disorder. The temperature at
which the maximum of that peak occurs is very close (but slightly
larger) to the one in the clean model.

In contrast to the XY model, the uniform susceptibility in the
presence of disorder increases rapidly with decreasing temperature at
the lowest temperatures accessible to us.  The susceptibility
therefore behaves qualitatively similar to the Ising model. It is
worth emphasizing that our results for all observables in the
honeycomb lattice appear to converge at temperatures significantly
below $T=0.1$, and quite close to $T=0.01$ for the energy, entropy,
and uniform susceptibility.

Our calculations for the Heisenberg model reach lower temperatures and
access regimes beyond what has been possible with quantum Monte Carlo
simulations. For example, Ref.~\onlinecite{sandvik94} studied the case
of diluting an antiferromagnetic model with ferromagnetic bonds of
varying concentration. The results presented there did not reach the
equal probability $J=\pm 1$ case discussed here because of the sign
problem. For lower concentrations of ferromagnetic bonds (below 30\%),
the calculations were still limited to temperatures above
$T\approx0.3$.

\section{Summary}
\label{sec:summary}

We have used numerical linked cluster expansions to study
thermodynamic properties of spin models with bimodal ($\pm J$) bond
disorder. The results reported are in the thermodynamic limit at
temperatures for which they are well converged.

We have unveiled various interesting effects of disorder in spin-1/2
Ising, XY, and Heisenberg models. For all models we find that disorder
leads to a saturation of the energy at the lowest temperatures
accessible to us, in a regime where the entropy is higher than in
clean systems. This makes apparent that there are many low lying
energy states. For the disordered classical Ising model, on both the
square and the honeycomb lattice, the divergence of the specific heat
in the clean case is replaced by a peak, and the uniform
susceptibility follows an inverse temperature law for all temperatures
in the presence of disorder. This was explicitly verified order by
order. In the Heisenberg model, we also find that the susceptibility
increases with decreasing temperature for all temperatures accessible
to us in the presence of disorder. In the XY model, on the other hand,
we find that the susceptibility exhibits a plateau at low
temperatures. On both the XY and Heisenberg models, our NLCE
calculations were able to resolve a peak in the specific heat.

\section{Acknowledgments}
We acknowledge support from the National Science Foundation (NSF),
Grant No. OCI-0904597, and the U.S.~Office of Naval Research.

\bibliography{disorder2d}

\begin{thebibliography}{55}%
\makeatletter
\providecommand \@ifxundefined [1]{%
 \@ifx{#1\undefined}
}%
\providecommand \@ifnum [1]{%
 \ifnum #1\expandafter \@firstoftwo
 \else \expandafter \@secondoftwo
 \fi
}%
\providecommand \@ifx [1]{%
 \ifx #1\expandafter \@firstoftwo
 \else \expandafter \@secondoftwo
 \fi
}%
\providecommand \natexlab [1]{#1}%
\providecommand \enquote  [1]{``#1''}%
\providecommand \bibnamefont  [1]{#1}%
\providecommand \bibfnamefont [1]{#1}%
\providecommand \citenamefont [1]{#1}%
\providecommand \href@noop [0]{\@secondoftwo}%
\providecommand \href [0]{\begingroup \@sanitize@url \@href}%
\providecommand \@href[1]{\@@startlink{#1}\@@href}%
\providecommand \@@href[1]{\endgroup#1\@@endlink}%
\providecommand \@sanitize@url [0]{\catcode `\\12\catcode `\$12\catcode
  `\&12\catcode `\#12\catcode `\^12\catcode `\_12\catcode `\%12\relax}%
\providecommand \@@startlink[1]{}%
\providecommand \@@endlink[0]{}%
\providecommand \url  [0]{\begingroup\@sanitize@url \@url }%
\providecommand \@url [1]{\endgroup\@href {#1}{\urlprefix }}%
\providecommand \urlprefix  [0]{URL }%
\providecommand \Eprint [0]{\href }%
\providecommand \doibase [0]{http://dx.doi.org/}%
\providecommand \selectlanguage [0]{\@gobble}%
\providecommand \bibinfo  [0]{\@secondoftwo}%
\providecommand \bibfield  [0]{\@secondoftwo}%
\providecommand \translation [1]{[#1]}%
\providecommand \BibitemOpen [0]{}%
\providecommand \bibitemStop [0]{}%
\providecommand \bibitemNoStop [0]{.\EOS\space}%
\providecommand \EOS [0]{\spacefactor3000\relax}%
\providecommand \BibitemShut  [1]{\csname bibitem#1\endcsname}%
\let\auto@bib@innerbib\@empty
\bibitem [{\citenamefont {Anderson}(1958)}]{anderson58}%
  \BibitemOpen
  \bibfield  {author} {\bibinfo {author} {\bibfnamefont {P.~W.}\ \bibnamefont
  {Anderson}},\ }\href {\doibase 10.1103/PhysRev.109.1492} {\bibfield
  {journal} {\bibinfo  {journal} {Phys. Rev.}\ }\textbf {\bibinfo {volume}
  {109}},\ \bibinfo {pages} {1492} (\bibinfo {year} {1958})}\BibitemShut
  {NoStop}%
\bibitem [{\citenamefont {Binder}\ and\ \citenamefont
  {Young}(1986)}]{binder86}%
  \BibitemOpen
  \bibfield  {author} {\bibinfo {author} {\bibfnamefont {K.}~\bibnamefont
  {Binder}}\ and\ \bibinfo {author} {\bibfnamefont {A.~P.}\ \bibnamefont
  {Young}},\ }\href {\doibase 10.1103/RevModPhys.58.801} {\bibfield  {journal}
  {\bibinfo  {journal} {Rev. Mod. Phys.}\ }\textbf {\bibinfo {volume} {58}},\
  \bibinfo {pages} {801} (\bibinfo {year} {1986})}\BibitemShut {NoStop}%
\bibitem [{\citenamefont {Weissman}(1993)}]{weissman93}%
  \BibitemOpen
  \bibfield  {author} {\bibinfo {author} {\bibfnamefont {M.}~\bibnamefont
  {Weissman}},\ }\href {\doibase 10.1103/RevModPhys.65.829} {\bibfield
  {journal} {\bibinfo  {journal} {Rev. Mod. Phys.}\ }\textbf {\bibinfo {volume}
  {65}},\ \bibinfo {pages} {829} (\bibinfo {year} {1993})}\BibitemShut
  {NoStop}%
\bibitem [{\citenamefont {Katzgraber}\ \emph {et~al.}(2007)\citenamefont
  {Katzgraber}, \citenamefont {Lee},\ and\ \citenamefont
  {Campbell}}]{Katzgraber07}%
  \BibitemOpen
  \bibfield  {author} {\bibinfo {author} {\bibfnamefont {H.~G.}\ \bibnamefont
  {Katzgraber}}, \bibinfo {author} {\bibfnamefont {L.~W.}\ \bibnamefont {Lee}},
  \ and\ \bibinfo {author} {\bibfnamefont {I.~A.}\ \bibnamefont {Campbell}},\
  }\href {\doibase 10.1103/PhysRevB.75.014412} {\bibfield  {journal} {\bibinfo
  {journal} {Phys. Rev. B}\ }\textbf {\bibinfo {volume} {75}},\ \bibinfo
  {pages} {014412} (\bibinfo {year} {2007})}\BibitemShut {NoStop}%
\bibitem [{\citenamefont {Parisen~Toldin}\ \emph {et~al.}(2011)\citenamefont
  {Parisen~Toldin}, \citenamefont {Pelissetto},\ and\ \citenamefont
  {Vicari}}]{parisen_francesco_11}%
  \BibitemOpen
  \bibfield  {author} {\bibinfo {author} {\bibfnamefont {F.}~\bibnamefont
  {Parisen~Toldin}}, \bibinfo {author} {\bibfnamefont {A.}~\bibnamefont
  {Pelissetto}}, \ and\ \bibinfo {author} {\bibfnamefont {E.}~\bibnamefont
  {Vicari}},\ }\href {\doibase 10.1103/PhysRevE.84.051116} {\bibfield
  {journal} {\bibinfo  {journal} {Phys. Rev. E}\ }\textbf {\bibinfo {volume}
  {84}},\ \bibinfo {pages} {051116} (\bibinfo {year} {2011})}\BibitemShut
  {NoStop}%
\bibitem [{\citenamefont {Thomas}\ and\ \citenamefont
  {Middleton}(2013)}]{thomas_middleton_13}%
  \BibitemOpen
  \bibfield  {author} {\bibinfo {author} {\bibfnamefont {C.}~\bibnamefont
  {Thomas}}\ and\ \bibinfo {author} {\bibfnamefont {A.}~\bibnamefont
  {Middleton}},\ }\href {\doibase 10.1103/PhysRevE.87.043303} {\bibfield
  {journal} {\bibinfo  {journal} {Phys. Rev. E}\ }\textbf {\bibinfo {volume}
  {87}},\ \bibinfo {pages} {043303} (\bibinfo {year} {2013})}\BibitemShut
  {NoStop}%
\bibitem [{\citenamefont {Thomas}\ \emph {et~al.}(2011)\citenamefont {Thomas},
  \citenamefont {Huse},\ and\ \citenamefont {Middleton}}]{thomas_huse_11}%
  \BibitemOpen
  \bibfield  {author} {\bibinfo {author} {\bibfnamefont {C.}~\bibnamefont
  {Thomas}}, \bibinfo {author} {\bibfnamefont {D.}~\bibnamefont {Huse}}, \ and\
  \bibinfo {author} {\bibfnamefont {A.}~\bibnamefont {Middleton}},\ }\href
  {\doibase 10.1103/PhysRevLett.107.047203} {\bibfield  {journal} {\bibinfo
  {journal} {Phys. Rev. Lett.}\ }\textbf {\bibinfo {volume} {107}},\ \bibinfo
  {pages} {047203} (\bibinfo {year} {2011})}\BibitemShut {NoStop}%
\bibitem [{\citenamefont {Henelius}\ and\ \citenamefont
  {Sandvik}(2000)}]{henelius00}%
  \BibitemOpen
  \bibfield  {author} {\bibinfo {author} {\bibfnamefont {P.}~\bibnamefont
  {Henelius}}\ and\ \bibinfo {author} {\bibfnamefont {A.}~\bibnamefont
  {Sandvik}},\ }\href {\doibase 10.1103/PhysRevB.62.1102} {\bibfield  {journal}
  {\bibinfo  {journal} {Phys. Rev. B}\ }\textbf {\bibinfo {volume} {62}},\
  \bibinfo {pages} {1102} (\bibinfo {year} {2000})}\BibitemShut {NoStop}%
\bibitem [{\citenamefont {Troyer}\ and\ \citenamefont
  {Wiese}(2005)}]{troyer05}%
  \BibitemOpen
  \bibfield  {author} {\bibinfo {author} {\bibfnamefont {M.}~\bibnamefont
  {Troyer}}\ and\ \bibinfo {author} {\bibfnamefont {U.-J.}\ \bibnamefont
  {Wiese}},\ }\href {\doibase 10.1103/PhysRevLett.94.170201} {\bibfield
  {journal} {\bibinfo  {journal} {Phys. Rev. Lett.}\ }\textbf {\bibinfo
  {volume} {94}},\ \bibinfo {pages} {170201} (\bibinfo {year}
  {2005})}\BibitemShut {NoStop}%
\bibitem [{\citenamefont {{Tang}}\ \emph {et~al.}()\citenamefont {{Tang}},
  \citenamefont {{Iyer}},\ and\ \citenamefont {{Rigol}}}]{tang2014}%
  \BibitemOpen
  \bibfield  {author} {\bibinfo {author} {\bibfnamefont {B.}~\bibnamefont
  {{Tang}}}, \bibinfo {author} {\bibfnamefont {D.}~\bibnamefont {{Iyer}}}, \
  and\ \bibinfo {author} {\bibfnamefont {M.}~\bibnamefont {{Rigol}}},\
  }\href@noop {} {\ }\Eprint {http://arxiv.org/abs/1411.0699} {arXiv:1411.0699}
  \BibitemShut {NoStop}%
\bibitem [{\citenamefont {Rigol}(2014{\natexlab{a}})}]{rigol_14a}%
  \BibitemOpen
  \bibfield  {author} {\bibinfo {author} {\bibfnamefont {M.}~\bibnamefont
  {Rigol}},\ }\href {\doibase 10.1103/PhysRevLett.112.170601} {\bibfield
  {journal} {\bibinfo  {journal} {Phys. Rev. Lett.}\ }\textbf {\bibinfo
  {volume} {112}},\ \bibinfo {pages} {170601} (\bibinfo {year}
  {2014}{\natexlab{a}})}\BibitemShut {NoStop}%
\bibitem [{\citenamefont {Wouters}\ \emph {et~al.}(2014)\citenamefont
  {Wouters}, \citenamefont {De~Nardis}, \citenamefont {Brockmann},
  \citenamefont {Fioretto}, \citenamefont {Rigol},\ and\ \citenamefont
  {Caux}}]{wouters_denardis_14}%
  \BibitemOpen
  \bibfield  {author} {\bibinfo {author} {\bibfnamefont {B.}~\bibnamefont
  {Wouters}}, \bibinfo {author} {\bibfnamefont {J.}~\bibnamefont {De~Nardis}},
  \bibinfo {author} {\bibfnamefont {M.}~\bibnamefont {Brockmann}}, \bibinfo
  {author} {\bibfnamefont {D.}~\bibnamefont {Fioretto}}, \bibinfo {author}
  {\bibfnamefont {M.}~\bibnamefont {Rigol}}, \ and\ \bibinfo {author}
  {\bibfnamefont {J.-S.}\ \bibnamefont {Caux}},\ }\href {\doibase
  10.1103/PhysRevLett.113.117202} {\bibfield  {journal} {\bibinfo  {journal}
  {Phys. Rev. Lett.}\ }\textbf {\bibinfo {volume} {113}},\ \bibinfo {pages}
  {117202} (\bibinfo {year} {2014})}\BibitemShut {NoStop}%
\bibitem [{\citenamefont {Rigol}(2014{\natexlab{b}})}]{rigol_14b}%
  \BibitemOpen
  \bibfield  {author} {\bibinfo {author} {\bibfnamefont {M.}~\bibnamefont
  {Rigol}},\ }\href {\doibase 10.1103/PhysRevE.90.031301} {\bibfield  {journal}
  {\bibinfo  {journal} {Phys. Rev. E}\ }\textbf {\bibinfo {volume} {90}},\
  \bibinfo {pages} {031301(R)} (\bibinfo {year}
  {2014}{\natexlab{b}})}\BibitemShut {NoStop}%
\bibitem [{\citenamefont {Sykes}\ and\ \citenamefont {Fisher}(1962)}]{sykes62}%
  \BibitemOpen
  \bibfield  {author} {\bibinfo {author} {\bibfnamefont {M.}~\bibnamefont
  {Sykes}}\ and\ \bibinfo {author} {\bibfnamefont {M.~E.}\ \bibnamefont
  {Fisher}},\ }\href {\doibase http://dx.doi.org/10.1016/0031-8914(62)90080-0}
  {\bibfield  {journal} {\bibinfo  {journal} {Physica}\ }\textbf {\bibinfo
  {volume} {28}},\ \bibinfo {pages} {919 } (\bibinfo {year}
  {1962})}\BibitemShut {NoStop}%
\bibitem [{\citenamefont {Weihong}\ \emph {et~al.}(1991)\citenamefont
  {Weihong}, \citenamefont {Oitmaa},\ and\ \citenamefont {Hamer}}]{weihong91}%
  \BibitemOpen
  \bibfield  {author} {\bibinfo {author} {\bibfnamefont {Z.}~\bibnamefont
  {Weihong}}, \bibinfo {author} {\bibfnamefont {J.}~\bibnamefont {Oitmaa}}, \
  and\ \bibinfo {author} {\bibfnamefont {C.}~\bibnamefont {Hamer}},\ }\href
  {\doibase 10.1103/PhysRevB.44.11869} {\bibfield  {journal} {\bibinfo
  {journal} {Phys. Rev. B}\ }\textbf {\bibinfo {volume} {44}},\ \bibinfo
  {pages} {11869} (\bibinfo {year} {1991})}\BibitemShut {NoStop}%
\bibitem [{\citenamefont {Di~Ciolo}\ \emph {et~al.}(2014)\citenamefont
  {Di~Ciolo}, \citenamefont {Carrasquilla}, \citenamefont {Becca},
  \citenamefont {Rigol},\ and\ \citenamefont
  {Galitski}}]{diciolo_carrasquilla_14}%
  \BibitemOpen
  \bibfield  {author} {\bibinfo {author} {\bibfnamefont {A.}~\bibnamefont
  {Di~Ciolo}}, \bibinfo {author} {\bibfnamefont {J.}~\bibnamefont
  {Carrasquilla}}, \bibinfo {author} {\bibfnamefont {F.}~\bibnamefont {Becca}},
  \bibinfo {author} {\bibfnamefont {M.}~\bibnamefont {Rigol}}, \ and\ \bibinfo
  {author} {\bibfnamefont {V.}~\bibnamefont {Galitski}},\ }\href {\doibase
  10.1103/PhysRevB.89.094413} {\bibfield  {journal} {\bibinfo  {journal} {Phys.
  Rev. B}\ }\textbf {\bibinfo {volume} {89}},\ \bibinfo {pages} {094413}
  (\bibinfo {year} {2014})}\BibitemShut {NoStop}%
\bibitem [{\citenamefont {Onsager}(1944)}]{Onsager_44}%
  \BibitemOpen
  \bibfield  {author} {\bibinfo {author} {\bibfnamefont {L.}~\bibnamefont
  {Onsager}},\ }\href {\doibase 10.1103/PhysRev.65.117} {\bibfield  {journal}
  {\bibinfo  {journal} {Phys. Rev.}\ }\textbf {\bibinfo {volume} {65}},\
  \bibinfo {pages} {117} (\bibinfo {year} {1944})}\BibitemShut {NoStop}%
\bibitem [{\citenamefont {Edwards}\ and\ \citenamefont
  {Anderson}(1975)}]{Edwards_75}%
  \BibitemOpen
  \bibfield  {author} {\bibinfo {author} {\bibfnamefont {S.~F.}\ \bibnamefont
  {Edwards}}\ and\ \bibinfo {author} {\bibfnamefont {P.~W.}\ \bibnamefont
  {Anderson}},\ }\href@noop {} {\bibfield  {journal} {\bibinfo  {journal}
  {J.~Phys.~F: Met.~Phys.}\ }\textbf {\bibinfo {volume} {5}},\ \bibinfo {pages}
  {965} (\bibinfo {year} {1975})}\BibitemShut {NoStop}%
\bibitem [{\citenamefont {Fisher}(1959)}]{fisher59}%
  \BibitemOpen
  \bibfield  {author} {\bibinfo {author} {\bibfnamefont {M.~E.}\ \bibnamefont
  {Fisher}},\ }\href {\doibase http://dx.doi.org/10.1016/S0031-8914(59)95411-4}
  {\bibfield  {journal} {\bibinfo  {journal} {Physica}\ }\textbf {\bibinfo
  {volume} {25}},\ \bibinfo {pages} {521 } (\bibinfo {year}
  {1959})}\BibitemShut {NoStop}%
\bibitem [{\citenamefont {Cho}\ and\ \citenamefont {Fisher}(1997)}]{cho97}%
  \BibitemOpen
  \bibfield  {author} {\bibinfo {author} {\bibfnamefont {S.}~\bibnamefont
  {Cho}}\ and\ \bibinfo {author} {\bibfnamefont {M.~P.~A.}\ \bibnamefont
  {Fisher}},\ }\href {\doibase 10.1103/PhysRevB.55.1025} {\bibfield  {journal}
  {\bibinfo  {journal} {Phys. Rev. B}\ }\textbf {\bibinfo {volume} {55}},\
  \bibinfo {pages} {1025} (\bibinfo {year} {1997})}\BibitemShut {NoStop}%
\bibitem [{\citenamefont {Blackman}\ \emph {et~al.}(1998)\citenamefont
  {Blackman}, \citenamefont {Gon\ifmmode~\mbox{\c{c}}\else \c{c}\fi{}alves},\
  and\ \citenamefont {Poulter}}]{Blackman1998}%
  \BibitemOpen
  \bibfield  {author} {\bibinfo {author} {\bibfnamefont {J.~A.}\ \bibnamefont
  {Blackman}}, \bibinfo {author} {\bibfnamefont {J.~R.}\ \bibnamefont
  {Gon\ifmmode~\mbox{\c{c}}\else \c{c}\fi{}alves}}, \ and\ \bibinfo {author}
  {\bibfnamefont {J.}~\bibnamefont {Poulter}},\ }\href {\doibase
  10.1103/PhysRevE.58.1502} {\bibfield  {journal} {\bibinfo  {journal} {Phys.
  Rev. E}\ }\textbf {\bibinfo {volume} {58}},\ \bibinfo {pages} {1502}
  (\bibinfo {year} {1998})}\BibitemShut {NoStop}%
\bibitem [{\citenamefont {Hartmann}\ and\ \citenamefont
  {Young}(2001)}]{hartman01}%
  \BibitemOpen
  \bibfield  {author} {\bibinfo {author} {\bibfnamefont {A.~K.}\ \bibnamefont
  {Hartmann}}\ and\ \bibinfo {author} {\bibfnamefont {A.~P.}\ \bibnamefont
  {Young}},\ }\href {\doibase 10.1103/PhysRevB.64.180404} {\bibfield  {journal}
  {\bibinfo  {journal} {Phys. Rev. B}\ }\textbf {\bibinfo {volume} {64}},\
  \bibinfo {pages} {180404} (\bibinfo {year} {2001})}\BibitemShut {NoStop}%
\bibitem [{\citenamefont {Gruzberg}\ \emph {et~al.}(2001)\citenamefont
  {Gruzberg}, \citenamefont {Read},\ and\ \citenamefont {Ludwig}}]{gruzberg01}%
  \BibitemOpen
  \bibfield  {author} {\bibinfo {author} {\bibfnamefont {I.}~\bibnamefont
  {Gruzberg}}, \bibinfo {author} {\bibfnamefont {N.}~\bibnamefont {Read}}, \
  and\ \bibinfo {author} {\bibfnamefont {A.}~\bibnamefont {Ludwig}},\ }\href
  {\doibase 10.1103/PhysRevB.63.104422} {\bibfield  {journal} {\bibinfo
  {journal} {Phys. Rev. B}\ }\textbf {\bibinfo {volume} {63}},\ \bibinfo
  {pages} {104422} (\bibinfo {year} {2001})}\BibitemShut {NoStop}%
\bibitem [{\citenamefont {Amoruso}\ and\ \citenamefont
  {Hartmann}(2004)}]{Amoruso_04}%
  \BibitemOpen
  \bibfield  {author} {\bibinfo {author} {\bibfnamefont {C.}~\bibnamefont
  {Amoruso}}\ and\ \bibinfo {author} {\bibfnamefont {A.~K.}\ \bibnamefont
  {Hartmann}},\ }\href {\doibase 10.1103/PhysRevB.70.134425} {\bibfield
  {journal} {\bibinfo  {journal} {Phys. Rev. B}\ }\textbf {\bibinfo {volume}
  {70}},\ \bibinfo {pages} {134425} (\bibinfo {year} {2004})}\BibitemShut
  {NoStop}%
\bibitem [{\citenamefont {Lukic}\ \emph {et~al.}(2004)\citenamefont {Lukic},
  \citenamefont {Galluccio}, \citenamefont {Marinari}, \citenamefont {Martin},\
  and\ \citenamefont {Rinaldi}}]{Lukic04}%
  \BibitemOpen
  \bibfield  {author} {\bibinfo {author} {\bibfnamefont {J.}~\bibnamefont
  {Lukic}}, \bibinfo {author} {\bibfnamefont {A.}~\bibnamefont {Galluccio}},
  \bibinfo {author} {\bibfnamefont {E.}~\bibnamefont {Marinari}}, \bibinfo
  {author} {\bibfnamefont {O.~C.}\ \bibnamefont {Martin}}, \ and\ \bibinfo
  {author} {\bibfnamefont {G.}~\bibnamefont {Rinaldi}},\ }\href {\doibase
  10.1103/PhysRevLett.92.117202} {\bibfield  {journal} {\bibinfo  {journal}
  {Phys. Rev. Lett.}\ }\textbf {\bibinfo {volume} {92}},\ \bibinfo {pages}
  {117202} (\bibinfo {year} {2004})}\BibitemShut {NoStop}%
\bibitem [{\citenamefont {Katzgraber}\ and\ \citenamefont
  {Lee}(2005)}]{Katzgraber05}%
  \BibitemOpen
  \bibfield  {author} {\bibinfo {author} {\bibfnamefont {H.~G.}\ \bibnamefont
  {Katzgraber}}\ and\ \bibinfo {author} {\bibfnamefont {L.~W.}\ \bibnamefont
  {Lee}},\ }\href {\doibase 10.1103/PhysRevB.71.134404} {\bibfield  {journal}
  {\bibinfo  {journal} {Phys. Rev. B}\ }\textbf {\bibinfo {volume} {71}},\
  \bibinfo {pages} {134404} (\bibinfo {year} {2005})}\BibitemShut {NoStop}%
\bibitem [{\citenamefont {Atisattapong}\ and\ \citenamefont
  {Poulter}(2009)}]{atisattapong09}%
  \BibitemOpen
  \bibfield  {author} {\bibinfo {author} {\bibfnamefont {W.}~\bibnamefont
  {Atisattapong}}\ and\ \bibinfo {author} {\bibfnamefont {J.}~\bibnamefont
  {Poulter}},\ }\href {http://stacks.iop.org/1367-2630/11/i=6/a=063039}
  {\bibfield  {journal} {\bibinfo  {journal} {New Journal of Physics}\ }\textbf
  {\bibinfo {volume} {11}},\ \bibinfo {pages} {063039} (\bibinfo {year}
  {2009})}\BibitemShut {NoStop}%
\bibitem [{\citenamefont {Mermin}\ and\ \citenamefont
  {Wagner}(1966)}]{mermin66}%
  \BibitemOpen
  \bibfield  {author} {\bibinfo {author} {\bibfnamefont {N.~D.}\ \bibnamefont
  {Mermin}}\ and\ \bibinfo {author} {\bibfnamefont {H.}~\bibnamefont
  {Wagner}},\ }\href {\doibase 10.1103/PhysRevLett.17.1133} {\bibfield
  {journal} {\bibinfo  {journal} {Phys. Rev. Lett.}\ }\textbf {\bibinfo
  {volume} {17}},\ \bibinfo {pages} {1133} (\bibinfo {year}
  {1966})}\BibitemShut {NoStop}%
\bibitem [{\citenamefont {{Altland}}\ and\ \citenamefont
  {{Simons}}(2006)}]{altland06}%
  \BibitemOpen
  \bibfield  {author} {\bibinfo {author} {\bibfnamefont {A.}~\bibnamefont
  {{Altland}}}\ and\ \bibinfo {author} {\bibfnamefont {B.}~\bibnamefont
  {{Simons}}},\ }\href {\doibase 10.2277/0521845084} {\emph {\bibinfo {title}
  {{Condensed Matter Field Theory}}}}\ (\bibinfo  {publisher} {Cambridge
  Univ.~Press},\ \bibinfo {year} {2006})\BibitemShut {NoStop}%
\bibitem [{\citenamefont {{Berezinski{\v i}}}(1972)}]{berezinski72}%
  \BibitemOpen
  \bibfield  {author} {\bibinfo {author} {\bibfnamefont {V.~L.}\ \bibnamefont
  {{Berezinski{\v i}}}},\ }\href@noop {} {\bibfield  {journal} {\bibinfo
  {journal} {Sov.~JETP}\ }\textbf {\bibinfo {volume} {34}},\ \bibinfo {pages}
  {610} (\bibinfo {year} {1972})}\BibitemShut {NoStop}%
\bibitem [{\citenamefont {Kosterlitz}\ and\ \citenamefont
  {Thouless}(1973)}]{kosterlitz73}%
  \BibitemOpen
  \bibfield  {author} {\bibinfo {author} {\bibfnamefont {J.~M.}\ \bibnamefont
  {Kosterlitz}}\ and\ \bibinfo {author} {\bibfnamefont {D.~J.}\ \bibnamefont
  {Thouless}},\ }\href {http://stacks.iop.org/0022-3719/6/i=7/a=010} {\bibfield
   {journal} {\bibinfo  {journal} {J.~Phys.~C: Solid State Phys.}\ }\textbf
  {\bibinfo {volume} {6}},\ \bibinfo {pages} {1181} (\bibinfo {year}
  {1973})}\BibitemShut {NoStop}%
\bibitem [{\citenamefont {{Harada}}\ and\ \citenamefont
  {{Kawashima}}(1998)}]{harada98}%
  \BibitemOpen
  \bibfield  {author} {\bibinfo {author} {\bibfnamefont {K.}~\bibnamefont
  {{Harada}}}\ and\ \bibinfo {author} {\bibfnamefont {N.}~\bibnamefont
  {{Kawashima}}},\ }\href {\doibase 10.1143/JPSJ.67.2768} {\bibfield  {journal}
  {\bibinfo  {journal} {J.~Phys.~Soc.~Jpn.}\ }\textbf {\bibinfo {volume}
  {67}},\ \bibinfo {pages} {2768} (\bibinfo {year} {1998})}\BibitemShut
  {NoStop}%
\bibitem [{\citenamefont {Carrasquilla}\ and\ \citenamefont
  {Rigol}(2012)}]{carrasquilla12}%
  \BibitemOpen
  \bibfield  {author} {\bibinfo {author} {\bibfnamefont {J.}~\bibnamefont
  {Carrasquilla}}\ and\ \bibinfo {author} {\bibfnamefont {M.}~\bibnamefont
  {Rigol}},\ }\href {\doibase 10.1103/PhysRevA.86.043629} {\bibfield  {journal}
  {\bibinfo  {journal} {Phys. Rev. A}\ }\textbf {\bibinfo {volume} {86}},\
  \bibinfo {pages} {043629} (\bibinfo {year} {2012})}\BibitemShut {NoStop}%
\bibitem [{\citenamefont {{Berezinski{\v i}}}(1971)}]{berezinskii71}%
  \BibitemOpen
  \bibfield  {author} {\bibinfo {author} {\bibfnamefont {V.~L.}\ \bibnamefont
  {{Berezinski{\v i}}}},\ }\href@noop {} {\bibfield  {journal} {\bibinfo
  {journal} {Sov.~JETP}\ }\textbf {\bibinfo {volume} {32}},\ \bibinfo {pages}
  {493} (\bibinfo {year} {1971})}\BibitemShut {NoStop}%
\bibitem [{\citenamefont {Fröhlich}\ and\ \citenamefont
  {Spencer}(1981)}]{frohlich81}%
  \BibitemOpen
  \bibfield  {author} {\bibinfo {author} {\bibfnamefont {J.}~\bibnamefont
  {Fröhlich}}\ and\ \bibinfo {author} {\bibfnamefont {T.}~\bibnamefont
  {Spencer}},\ }\href {\doibase 10.1007/BF01208273} {\bibfield  {journal}
  {\bibinfo  {journal} {Comm.~Math.~Phys.}\ }\textbf {\bibinfo {volume} {81}},\
  \bibinfo {pages} {527} (\bibinfo {year} {1981})}\BibitemShut {NoStop}%
\bibitem [{\citenamefont {{Kenna}}(2005)}]{kenna05}%
  \BibitemOpen
  \bibfield  {author} {\bibinfo {author} {\bibfnamefont {R.}~\bibnamefont
  {{Kenna}}},\ }\href@noop {} {\bibfield  {journal} {\bibinfo  {journal}
  {arXiv:cond-mat/0512356}\ } (\bibinfo {year} {2005})}\BibitemShut {NoStop}%
\bibitem [{\citenamefont {Gross}\ and\ \citenamefont
  {Wilczek}(1973)}]{gross73}%
  \BibitemOpen
  \bibfield  {author} {\bibinfo {author} {\bibfnamefont {D.}~\bibnamefont
  {Gross}}\ and\ \bibinfo {author} {\bibfnamefont {F.}~\bibnamefont
  {Wilczek}},\ }\href {\doibase 10.1103/PhysRevLett.30.1343} {\bibfield
  {journal} {\bibinfo  {journal} {Phys. Rev. Lett.}\ }\textbf {\bibinfo
  {volume} {30}},\ \bibinfo {pages} {1343} (\bibinfo {year}
  {1973})}\BibitemShut {NoStop}%
\bibitem [{\citenamefont {Gross}(1999)}]{gross99}%
  \BibitemOpen
  \bibfield  {author} {\bibinfo {author} {\bibfnamefont {D.~J.}\ \bibnamefont
  {Gross}},\ }\href {\doibase http://dx.doi.org/10.1016/S0920-5632(99)00208-X}
  {\bibfield  {journal} {\bibinfo  {journal} {Nucl.~Phys.~B - Proc.~Suppl.}\
  }\textbf {\bibinfo {volume} {74}},\ \bibinfo {pages} {426 } (\bibinfo {year}
  {1999})},\ \bibinfo {note} {\{QCD\} 98}\BibitemShut {NoStop}%
\bibitem [{\citenamefont {{Cuccoli}}\ \emph {et~al.}(1997)\citenamefont
  {{Cuccoli}}, \citenamefont {{Tognetti}}, \citenamefont {{Vaia}},\ and\
  \citenamefont {{Verrucchi}}}]{cuccoli97}%
  \BibitemOpen
  \bibfield  {author} {\bibinfo {author} {\bibfnamefont {A.}~\bibnamefont
  {{Cuccoli}}}, \bibinfo {author} {\bibfnamefont {V.}~\bibnamefont
  {{Tognetti}}}, \bibinfo {author} {\bibfnamefont {R.}~\bibnamefont {{Vaia}}},
  \ and\ \bibinfo {author} {\bibfnamefont {P.}~\bibnamefont {{Verrucchi}}},\
  }\href {\doibase 10.1103/PhysRevB.56.14456} {\bibfield  {journal} {\bibinfo
  {journal} {\prb}\ }\textbf {\bibinfo {volume} {56}},\ \bibinfo {pages}
  {14456} (\bibinfo {year} {1997})}\BibitemShut {NoStop}%
\bibitem [{\citenamefont {Domb}(1960{\natexlab{a}})}]{domb60a}%
  \BibitemOpen
  \bibfield  {author} {\bibinfo {author} {\bibfnamefont {C.}~\bibnamefont
  {Domb}},\ }\href {\doibase 10.1080/00018736000101189} {\bibfield  {journal}
  {\bibinfo  {journal} {Advances in Physics}\ }\textbf {\bibinfo {volume}
  {9}},\ \bibinfo {pages} {149} (\bibinfo {year} {1960}{\natexlab{a}})},\
  \Eprint {http://arxiv.org/abs/http://dx.doi.org/10.1080/00018736000101189}
  {http://dx.doi.org/10.1080/00018736000101189} \BibitemShut {NoStop}%
\bibitem [{\citenamefont {Domb}(1960{\natexlab{b}})}]{domb60b}%
  \BibitemOpen
  \bibfield  {author} {\bibinfo {author} {\bibfnamefont {C.}~\bibnamefont
  {Domb}},\ }\href {\doibase 10.1080/00018736000101199} {\bibfield  {journal}
  {\bibinfo  {journal} {Advances in Physics}\ }\textbf {\bibinfo {volume}
  {9}},\ \bibinfo {pages} {245} (\bibinfo {year} {1960}{\natexlab{b}})},\
  \Eprint {http://arxiv.org/abs/http://dx.doi.org/10.1080/00018736000101199}
  {http://dx.doi.org/10.1080/00018736000101199} \BibitemShut {NoStop}%
\bibitem [{\citenamefont {{Sykes}}\ \emph {et~al.}(1966)\citenamefont
  {{Sykes}}, \citenamefont {{Essam}}, \citenamefont {{Heap}},\ and\
  \citenamefont {{Hiley}}}]{sykes66}%
  \BibitemOpen
  \bibfield  {author} {\bibinfo {author} {\bibfnamefont {M.~F.}\ \bibnamefont
  {{Sykes}}}, \bibinfo {author} {\bibfnamefont {J.~W.}\ \bibnamefont
  {{Essam}}}, \bibinfo {author} {\bibfnamefont {B.~R.}\ \bibnamefont {{Heap}}},
  \ and\ \bibinfo {author} {\bibfnamefont {B.~J.}\ \bibnamefont {{Hiley}}},\
  }\href {\doibase 10.1063/1.1705066} {\bibfield  {journal} {\bibinfo
  {journal} {Journal of Mathematical Physics}\ }\textbf {\bibinfo {volume}
  {7}},\ \bibinfo {pages} {1557} (\bibinfo {year} {1966})}\BibitemShut
  {NoStop}%
\bibitem [{\citenamefont {Rigol}\ \emph {et~al.}(2006)\citenamefont {Rigol},
  \citenamefont {Bryant},\ and\ \citenamefont {Singh}}]{rigol_bryant_06}%
  \BibitemOpen
  \bibfield  {author} {\bibinfo {author} {\bibfnamefont {M.}~\bibnamefont
  {Rigol}}, \bibinfo {author} {\bibfnamefont {T.}~\bibnamefont {Bryant}}, \
  and\ \bibinfo {author} {\bibfnamefont {R.~R.~P.}\ \bibnamefont {Singh}},\
  }\href {\doibase 10.1103/PhysRevLett.97.187202} {\bibfield  {journal}
  {\bibinfo  {journal} {Phys. Rev. Lett.}\ }\textbf {\bibinfo {volume} {97}},\
  \bibinfo {pages} {187202} (\bibinfo {year} {2006})}\BibitemShut {NoStop}%
\bibitem [{\citenamefont {Rigol}\ \emph
  {et~al.}(2007{\natexlab{a}})\citenamefont {Rigol}, \citenamefont {Bryant},\
  and\ \citenamefont {Singh}}]{rigol_bryant_07}%
  \BibitemOpen
  \bibfield  {author} {\bibinfo {author} {\bibfnamefont {M.}~\bibnamefont
  {Rigol}}, \bibinfo {author} {\bibfnamefont {T.}~\bibnamefont {Bryant}}, \
  and\ \bibinfo {author} {\bibfnamefont {R.~R.~P.}\ \bibnamefont {Singh}},\
  }\href {\doibase 10.1103/PhysRevE.75.061118} {\bibfield  {journal} {\bibinfo
  {journal} {Phys. Rev. E}\ }\textbf {\bibinfo {volume} {75}},\ \bibinfo
  {pages} {061118} (\bibinfo {year} {2007}{\natexlab{a}})}\BibitemShut
  {NoStop}%
\bibitem [{\citenamefont {Rigol}\ \emph
  {et~al.}(2007{\natexlab{b}})\citenamefont {Rigol}, \citenamefont {Bryant},\
  and\ \citenamefont {Singh}}]{rigol_bryant_07b}%
  \BibitemOpen
  \bibfield  {author} {\bibinfo {author} {\bibfnamefont {M.}~\bibnamefont
  {Rigol}}, \bibinfo {author} {\bibfnamefont {T.}~\bibnamefont {Bryant}}, \
  and\ \bibinfo {author} {\bibfnamefont {R.~R.~P.}\ \bibnamefont {Singh}},\
  }\href {\doibase 10.1103/PhysRevE.75.061119} {\bibfield  {journal} {\bibinfo
  {journal} {Phys. Rev. E}\ }\textbf {\bibinfo {volume} {75}},\ \bibinfo
  {pages} {061119} (\bibinfo {year} {2007}{\natexlab{b}})}\BibitemShut
  {NoStop}%
\bibitem [{\citenamefont {{Khatami}}\ and\ \citenamefont
  {{Rigol}}(2011{\natexlab{a}})}]{khatami11a}%
  \BibitemOpen
  \bibfield  {author} {\bibinfo {author} {\bibfnamefont {E.}~\bibnamefont
  {{Khatami}}}\ and\ \bibinfo {author} {\bibfnamefont {M.}~\bibnamefont
  {{Rigol}}},\ }\href {\doibase 10.1103/PhysRevB.83.134431} {\bibfield
  {journal} {\bibinfo  {journal} {\prb}\ }\textbf {\bibinfo {volume} {83}},\
  \bibinfo {eid} {134431} (\bibinfo {year} {2011}{\natexlab{a}})}\BibitemShut
  {NoStop}%
\bibitem [{\citenamefont {{Khatami}}\ and\ \citenamefont
  {{Rigol}}(2011{\natexlab{b}})}]{khatami11b}%
  \BibitemOpen
  \bibfield  {author} {\bibinfo {author} {\bibfnamefont {E.}~\bibnamefont
  {{Khatami}}}\ and\ \bibinfo {author} {\bibfnamefont {M.}~\bibnamefont
  {{Rigol}}},\ }\href {\doibase 10.1103/PhysRevA.84.053611} {\bibfield
  {journal} {\bibinfo  {journal} {\pra}\ }\textbf {\bibinfo {volume} {84}},\
  \bibinfo {eid} {053611} (\bibinfo {year} {2011}{\natexlab{b}})}\BibitemShut
  {NoStop}%
\bibitem [{\citenamefont {{Tang}}\ \emph {et~al.}(2012)\citenamefont {{Tang}},
  \citenamefont {{Paiva}}, \citenamefont {{Khatami}},\ and\ \citenamefont
  {{Rigol}}}]{tang12}%
  \BibitemOpen
  \bibfield  {author} {\bibinfo {author} {\bibfnamefont {B.}~\bibnamefont
  {{Tang}}}, \bibinfo {author} {\bibfnamefont {T.}~\bibnamefont {{Paiva}}},
  \bibinfo {author} {\bibfnamefont {E.}~\bibnamefont {{Khatami}}}, \ and\
  \bibinfo {author} {\bibfnamefont {M.}~\bibnamefont {{Rigol}}},\ }\href
  {\doibase 10.1103/PhysRevLett.109.205301} {\bibfield  {journal} {\bibinfo
  {journal} {Phys.~Rev.~Lett.}\ }\textbf {\bibinfo {volume} {109}},\ \bibinfo
  {eid} {205301} (\bibinfo {year} {2012})}\BibitemShut {NoStop}%
\bibitem [{\citenamefont {{Tang}}\ \emph {et~al.}(2013)\citenamefont {{Tang}},
  \citenamefont {{Paiva}}, \citenamefont {{Khatami}},\ and\ \citenamefont
  {{Rigol}}}]{tang13b}%
  \BibitemOpen
  \bibfield  {author} {\bibinfo {author} {\bibfnamefont {B.}~\bibnamefont
  {{Tang}}}, \bibinfo {author} {\bibfnamefont {T.}~\bibnamefont {{Paiva}}},
  \bibinfo {author} {\bibfnamefont {E.}~\bibnamefont {{Khatami}}}, \ and\
  \bibinfo {author} {\bibfnamefont {M.}~\bibnamefont {{Rigol}}},\ }\href
  {\doibase 10.1103/PhysRevB.88.125127} {\bibfield  {journal} {\bibinfo
  {journal} {\prb}\ }\textbf {\bibinfo {volume} {88}},\ \bibinfo {eid} {125127}
  (\bibinfo {year} {2013})}\BibitemShut {NoStop}%
\bibitem [{\citenamefont {Tang}\ \emph {et~al.}(2013)\citenamefont {Tang},
  \citenamefont {Khatami},\ and\ \citenamefont {Rigol}}]{tang_khatami_13}%
  \BibitemOpen
  \bibfield  {author} {\bibinfo {author} {\bibfnamefont {B.}~\bibnamefont
  {Tang}}, \bibinfo {author} {\bibfnamefont {E.}~\bibnamefont {Khatami}}, \
  and\ \bibinfo {author} {\bibfnamefont {M.}~\bibnamefont {Rigol}},\
  }\href@noop {} {\bibfield  {journal} {\bibinfo  {journal} {Comput. Phys.
  Commun.}\ }\textbf {\bibinfo {volume} {184}},\ \bibinfo {pages} {557}
  (\bibinfo {year} {2013})}\BibitemShut {NoStop}%
\bibitem [{\citenamefont {Khatami}\ \emph {et~al.}(2011)\citenamefont
  {Khatami}, \citenamefont {Singh},\ and\ \citenamefont
  {Rigol}}]{khatami_singh_11}%
  \BibitemOpen
  \bibfield  {author} {\bibinfo {author} {\bibfnamefont {E.}~\bibnamefont
  {Khatami}}, \bibinfo {author} {\bibfnamefont {R.~R.~P.}\ \bibnamefont
  {Singh}}, \ and\ \bibinfo {author} {\bibfnamefont {M.}~\bibnamefont
  {Rigol}},\ }\href {\doibase 10.1103/PhysRevB.84.224411} {\bibfield  {journal}
  {\bibinfo  {journal} {Phys. Rev. B}\ }\textbf {\bibinfo {volume} {84}},\
  \bibinfo {pages} {224411} (\bibinfo {year} {2011})}\BibitemShut {NoStop}%
\bibitem [{\citenamefont {Huang}(1987)}]{huang87}%
  \BibitemOpen
  \bibfield  {author} {\bibinfo {author} {\bibfnamefont {K.}~\bibnamefont
  {Huang}},\ }\href {http://books.google.com/books?id=M8PvAAAAMAAJ} {\emph
  {\bibinfo {title} {Statistical mechanics}}}\ (\bibinfo  {publisher} {Wiley},\
  \bibinfo {address} {New York},\ \bibinfo {year} {1987})\BibitemShut {NoStop}%
\bibitem [{\citenamefont {{Deng}}\ and\ \citenamefont {{Gu}}(2014)}]{deng14}%
  \BibitemOpen
  \bibfield  {author} {\bibinfo {author} {\bibfnamefont {Y.-B.}\ \bibnamefont
  {{Deng}}}\ and\ \bibinfo {author} {\bibfnamefont {Q.}~\bibnamefont {{Gu}}},\
  }\href {\doibase 10.1088/0256-307X/31/2/020504} {\bibfield  {journal}
  {\bibinfo  {journal} {Chin.~Phys.~Lett.}\ }\textbf {\bibinfo {volume} {31}},\
  \bibinfo {eid} {020504} (\bibinfo {year} {2014})}\BibitemShut {NoStop}%
\bibitem [{\citenamefont {{Surungan}}\ and\ \citenamefont
  {{Okabe}}(2005)}]{surungan05}%
  \BibitemOpen
  \bibfield  {author} {\bibinfo {author} {\bibfnamefont {T.}~\bibnamefont
  {{Surungan}}}\ and\ \bibinfo {author} {\bibfnamefont {Y.}~\bibnamefont
  {{Okabe}}},\ }\href {\doibase 10.1103/PhysRevB.71.184438} {\bibfield
  {journal} {\bibinfo  {journal} {\prb}\ }\textbf {\bibinfo {volume} {71}},\
  \bibinfo {eid} {184438} (\bibinfo {year} {2005})}\BibitemShut {NoStop}%
\bibitem [{\citenamefont {Sandvik}(1994)}]{sandvik94}%
  \BibitemOpen
  \bibfield  {author} {\bibinfo {author} {\bibfnamefont {A.~W.}\ \bibnamefont
  {Sandvik}},\ }\href {\doibase 10.1103/PhysRevB.50.15803} {\bibfield
  {journal} {\bibinfo  {journal} {Phys. Rev. B}\ }\textbf {\bibinfo {volume}
  {50}},\ \bibinfo {pages} {15803} (\bibinfo {year} {1994})}\BibitemShut
  {NoStop}%
\end{thebibliography}%
\end{document}